\documentclass[authoryear,12pt]{elsarticle}

\usepackage{amssymb}

\usepackage{amsmath}
\usepackage{endnotes}

\def\beq{\begin{equation}}
\def\eeq#1{\label{#1}\end{equation}}
\def\beqa{\begin{eqnarray}}
\def\eeqa#1{\label{#1}\end{eqnarray}}

\def\bigO{{\cal O}}

\def\sign{{\rm sign}}

\journal{Journal of Theoretical Biology}

\begin{document}

\vskip-10cm
\begin{frontmatter}



\title{A sampling theory for asymmetric communities}


\author[one]{Andrew E. Noble\corref{cor1}}
\ead{andrewenoble@gmail.com}

\author[two]{Nico M. Temme}
\ead{Nico.Temme@cwi.nl}

\author[one]{William F. Fagan}
\ead{bfagan@umd.edu}

\author[three]{Timothy H. Keitt}
\ead{tkeitt@gmail.com}

\cortext[cor1]{Corresponding author}

\address[one]{Department of Biology, University of Maryland, College Park, MD 20742, USA}
\address[two]{Center for Mathematics and Computer Science, Science Park 123, 1098 XG Amsterdam, The Netherlands}
\address[three]{Section of Integrative Biology, University of Texas at Austin, 1 University Station C0930, Austin, TX 78712, USA}

\begin{abstract}
We introduce the first analytical model of asymmetric community dynamics to yield Hubbell's neutral theory in the limit of functional equivalence among all species.  Our focus centers on an asymmetric extension of Hubbell's local community dynamics, while an analogous extension of Hubbell's metacommunity dynamics is deferred to an appendix.  We find that mass-effects may facilitate coexistence in asymmetric local communities and generate unimodal species abundance distributions indistinguishable from those of symmetric communities.  Multiple modes, however, only arise from asymmetric processes and provide a strong indication of non-neutral dynamics.  Although the exact stationary distributions of fully asymmetric communities must be calculated numerically, we derive approximate sampling distributions for the general case and for nearly neutral communities where symmetry is broken by a single species distinct from all others in ecological fitness and dispersal ability.  In the latter case, our approximate distributions are fully normalized, and novel asymptotic expansions of the required hypergeometric functions are provided to make evaluations tractable for large communities.  Employing these results in a Bayesian analysis may provide a novel statistical test to assess the consistency of species abundance data with the neutral hypothesis.
\end{abstract}


\end{frontmatter}


\section{Introduction}
\label{intro}
The ecological symmetry of trophically similar species forms the central assumption in Hubbell's unified neutral theory of biodiversity and biogeography~\citep{PHubbell:2001p4284}.  In the absence of stable coexistence mechanisms, local communities evolve under zero-sum ecological drift -- a stochastic process of density-dependent birth, death, and migration that maintains a fixed community size~\citep{PHubbell:2001p4284}.  Despite a homogeneous environment, migration inhibits the dominance of any single species and fosters high levels of diversity.  The symmetry assumption has allowed for considerable analytical developments that draw on the mathematics of neutral population genetics~\citep{Fisher:1930,Wright:1931} to derive exact predictions for emergent, macro-ecological patterns~\citep{Chave:2004p1260,Etienne:2007p3828,McKane:2000p4318,Vallade:2003p3851,Volkov:2003p1299,Etienne:2004p5302,McKane:2004p5661,Pigolotti:2004p5308,He:2005p7106,Volkov:2005p6326,Hu:2007p7107,Volkov:2007p1300,Babak:2008p7108,Babak:2009p7109}.  Among the most significant contributions are calculations of multivariate sampling distributions that relate local abundances to those in the regional metacommunity~\citep{Alonso:2004p4321,Etienne:2005p3829,Etienne:2005p6998,Etienne:2007p7111}.  Hubbell first emphasized the ultility of sampling theories for testing neutral theory against observed species abundance distributions (SADs)~\citep{PHubbell:2001p4284}.  Since then, Etienne and Olff have incorporated sampling distributions as conditional likelihoods in Bayesian analyses~\citep{Etienne:2004p5302,Etienne:2005p6363,Etienne:2007p7111,Etienne:2009p8436}.  Recent work has shown that the sampling distributions of neutral theory remain invariant when the restriction of zero-sum dynamics is lifted~\citep{Etienne:2007p5640,Haegeman:2008p6364,Conlisk:2010p8340} and when the assumption of strict symmetry is relaxed to a requirement of ecological equivalence~\citep{Etienne:2007p5640,Haegeman:2008p6364,Allouche:2009p6354,Allouche:2009p6355,Lin:2009p6368}.

The success of neutral theory in fitting empirical patterns of biodiversity~\citep{PHubbell:2001p4284,Volkov:2003p1299,Volkov:2005p6326,He:2005p7106,Chave:2006p6372} has generated a heated debate among ecologists, as there is strong evidence for species asymmetry in the field~\citep{Harper:1977,Goldberg:1992p6747,Chase:2003, Wootton:2005, Levine:2009p6069}.  Echoing previous work on the difficulty of resolving competitive dynamics from the essentially static observations of co-occurence data~\citep{Hastings:1987p8336}, recent studies indicate that interspecific tradeoffs may generate unimodal SADs indistinguishable from the expectations of neutral theory~\citep{Chave:2002p6357,Mouquet:2003p6380,Chase:2005p6356,He:2005p7106,Purves:2005,Walker:2007p4670,Doncaster:2009p6361}.  These results underscore the compatibility of asymmetries and coexistence.  The pioneering work of~\citet{Hutchinson:1951p7110}, has inspired a large literature on asymmetries in dispersal ability that permit the coexistence of ``fugitive species" with dominant competitors.  In particular, \citet{Shmida:1985p7047} extended the work of \citet{Brown:1977p7046} by introducing the paradigm of ``mass-effects", where immigration facilitates the establishment of species in sites where they would otherwise be competitively excluded.  Numerous attempts have been made to reconcile such deterministic approaches to the coexistence of asymmetric species with the stochastic model of ecological drift in symmetric neutral theory~\citep{Zhang:1997p6376,Tilman:2004p4673,Chase:2005p6356,Alonso:2006p4856,Gravel:2006p4668, Pueyo:2007p1298, Walker:2007p4670, Alonso:2008p4724,Ernest:2008p1316,Zhou:2008p6377}.  Many of these attempts build on insights from the concluding chapter of Hubbell's book~\citep{PHubbell:2001p4284}.

Nevertheless, the need remains for a fully asymmetric, analytical, sampling theory that contains Hubbell's model as a limiting case~\citep{Alonso:2006p4856}.  In this article, we develop such a theory for local, dispersal-limited communities in the main text and defer an analogous treatment of metacommunities to Appendix A.  Hubbell's assumption of zero-sum dynamics is preserved, but the requirement of per capita ecological equivalence among all species is eliminated.  Asymmetries are introduced by allowing for the variations in ecological fitness and dispersal ability that may arise in a heterogeneous environment~\citep{Leibold:2004p1285,Holyoak:2005}.  Our work expands on the numerical simulations of~\citet{Zhou:2008p6377}, where variations in ecological fitness alone were considered.  Coexistence emerges from mass-effects as well as ecological equivalence, and both mechanisms generate unimodal SADs that may be indistinguishable.  For local communities and metacommunities, we derive approximate sampling distributions for both the general case and the nearly neutral case, where symmetry is broken by a single species unique in ecological function.  These approximations yield the sampling distributions of Hubbell's neutral model in the limit of functional equivalence among all species.  

\section{A general sampling theory for local communities}
\label{gen}
For a local community of $J_L$ individuals and $S$ possible species, we model community dynamics as a stochastic process, $\vec{N}(\tau)$, over the labelled community abundance vectors $\vec{n}=(n_1,\dots,n_S)$.  Consistent with zero-sum dynamics, we require all accessible states to contain $J_L$ total individuals:  $\sum_{i=1}^S n_i=J_L$ and $0\leq n_i \leq J_L$ .  The number of accessible states is $A=\sum_{n_1=0}^{J_L} \dots \sum_{n_S=0}^{J_L}$ $\delta(J_L-n_1-\dots-n_S)$.  

Allowed transitions first remove an individual from species $i$ and then add an individual to species $j$.  Removals are due to death or emigration and occur with the density-dependent probability $n_i/J_L$.  Additions are due either to an immigration event, with probability $m_j$, or a birth event, with probability $1-m_j$.  We will refer to the $m_j$ as dispersal abilities.  If immigration occurs, we assume that metacommunity relative abundance, $x_j$, determines the proportional representation of species $j$ in the propagule rain and that the probability of establishment is weighted by ecological fitness, $w_j$, where high values correspond to a local competitive advantage or a superior adaptation to the local environment.  Therefore, species $j$ recruits with probability
\beq
\frac{w_jx_j}{\sum_{k=1}^Sw_kx_k},
\eeq{othbirthprob}
where $x_j\in(0,1)$, $w_j\in(0,\infty)$, and $\sum_{k=1}^Sx_k=1$.  If immigration does not occur, we assume that local relative abundance, $n_j/J_L$, governs propagule rain composition such that species $j$ recruits with probability
\beq
\frac{w_jn_j}{\sum_{k=1}^Sw_kn_k-w_i}.
\eeq{birthprob}
In numerical simulations of an asymmetric community, \citet{Zhou:2008p6377} employed a similar probability for recruitment in the absence of immigration.  Here, a factor of $w_i$ is subtracted in the denominator because species $i$ loses an individual prior to the birth event for species $j$.  An analogous subtraction is absent from Eq.~\ref{othbirthprob} because we assume an infinite metacommunity where the $x_j$ are invariant to fluctuations in the finite, local community populations.       

In sum, the nonzero transition probabilities are stationary and given by
\beqa
T_{ij\vec{n}} &=&\lim_{\Delta \tau\to0}\frac{Pr\{\vec{N}(\tau+\Delta \tau)=\vec{n}-\vec{e}_i+\vec{e}_j | \vec{N}(\tau)=\vec{n}\}}{\Delta \tau}\nonumber \\
&=& \frac{n_i}{J_L}\left((1-m_j)\frac{w_jn_j}{\sum_{k=1}^Sw_kn_k-w_i}+m_j\frac{w_jx_j}{\sum_{k=1}^Sw_kx_k}\right),
\eeqa{t}
where $\vec{e}_i$ is an $S$--dimensional unit vector along the $i$th--direction, the $w_k$ must be sufficiently large such that $\sum_{k=1}^Sw_kn_k-w_i>0$, and the time, $\tau$, is dimensionless with a scale set by the overall transition rate.  The probability of state occupancy, $P_{\vec{n}}$, evolves according to the master equation
\beq
\frac{dP_{\vec{n}}}{d\tau}\,=\,\sum_{i=1}^S\sum_{j=1,j\ne i}^S\left(T_{ij\vec{n}+\vec{e}_i-\vec{e}_j}P_{\vec{n}+\vec{e}_i-\vec{e}_j}-T_{ji\vec{n}}P_{\vec{n}} \right) \Theta_{ij},
\eeq{evolve1}
where
\beq
\Theta_{ij}\,=\,\Theta(J_L-(n_i+1))\Theta(n_j-1),
\eeq{Theta}
and we define the step-function $\Theta(x)$ to be zero for $x<0$ and one otherwise.  Eq.~\ref{evolve1} can be recast in terms of a transition probability matrix $W$
\beq
\frac{dP_a}{d\tau}=\sum_{b=1}^AP_{b} W_{ba},
\eeq{evolve2}
where $a,b\in(1, \dots, A)$ enumerate accessible states with components $(a_1, \dots,$ $a_S)$, $(b_1, \dots,$ $b_S)$.  The left eigenvector of $W$ with zero eigenvalue yields the stationary distribution for community composition, $P_a^*\equiv\lim_{\tau\to\infty}P_a(\tau)$.  Marginal distributions yield the equilibrium abundance probabilities for each species $i$
\beq
P^{(i)*}_{n}=\sum_{a=1}^{A} \delta_{a_i,n} P_a^*.
\eeq{marg}
From here, we calculate the stationary SAD by following the general treatment of asymmetric communities in~\citet{Alonso:2008p4724}
\beq
S^*_n\,=\,\sum_{i=1}^SP^{(i)*}_{n}.
\eeq{SAD}
The expected species richness is
\beq
S^*\,=\,\sum_{n=1}^SS_n^*<S.
\eeq{Star}
Given that the local community, with abundances $n_i$, is defined as a sample of the metacommunity, with relative abundances $x_i$,  we have established the framework for a general sampling theory of local communities.

This sampling theory incorporates aspects of the mass-effects paradigm \citep{Brown:1977p7046,Shmida:1985p7047,Holt:1993p7063,Leibold:2004p1285,Holyoak:2005}.  Local asymmetries in ecological fitness imply environmental heterogeneity across the metacommunity such that competitive ability peaks in the local communities where biotic and abiotic factors most closely match niche requirements~\citep{Tilman:1982p7101,Leibold:1998p7099,Chase:2003}.  Where species experience a competitive disadvantage, the mass-effects of immigration allow for persistence.  Indeed, the master equation given by Eq.~\ref{evolve1}, when applied to open communities where $m_j>0$ for all $j$, admits no absorbing states and ensures that every species has a nonzero probability of being present under equilibrium conditions.  By contrast, when Eq.~\ref{evolve1} is applied to closed communities where $m_j=0$ for all $j$, the eventual dominance of a single species is guaranteed.  

\begin{figure*}
\vskip-2cm
\centerline{
\includegraphics[width=.602\textwidth]{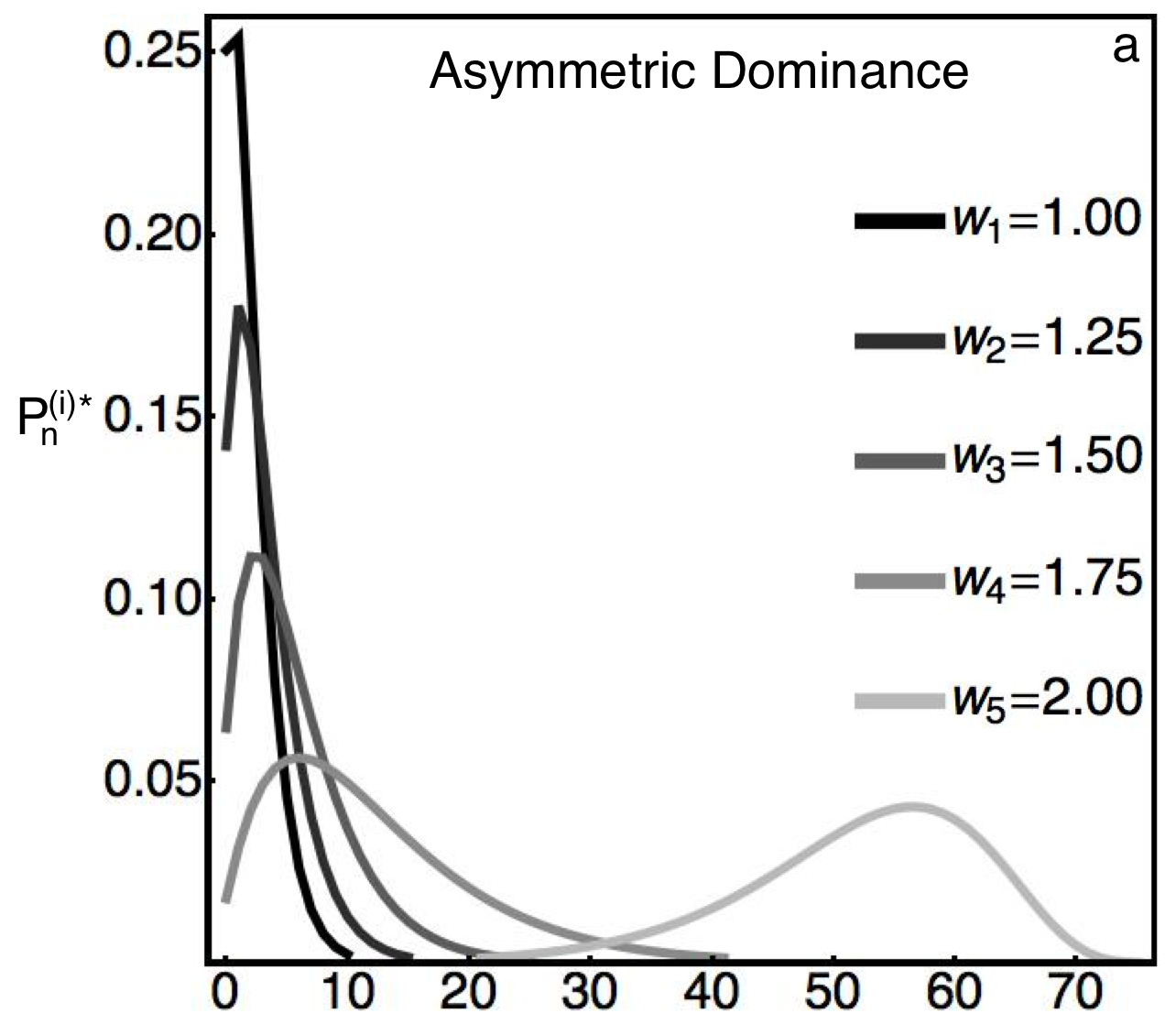}
\hskip-0.27cm
\includegraphics[width=.5067\textwidth]{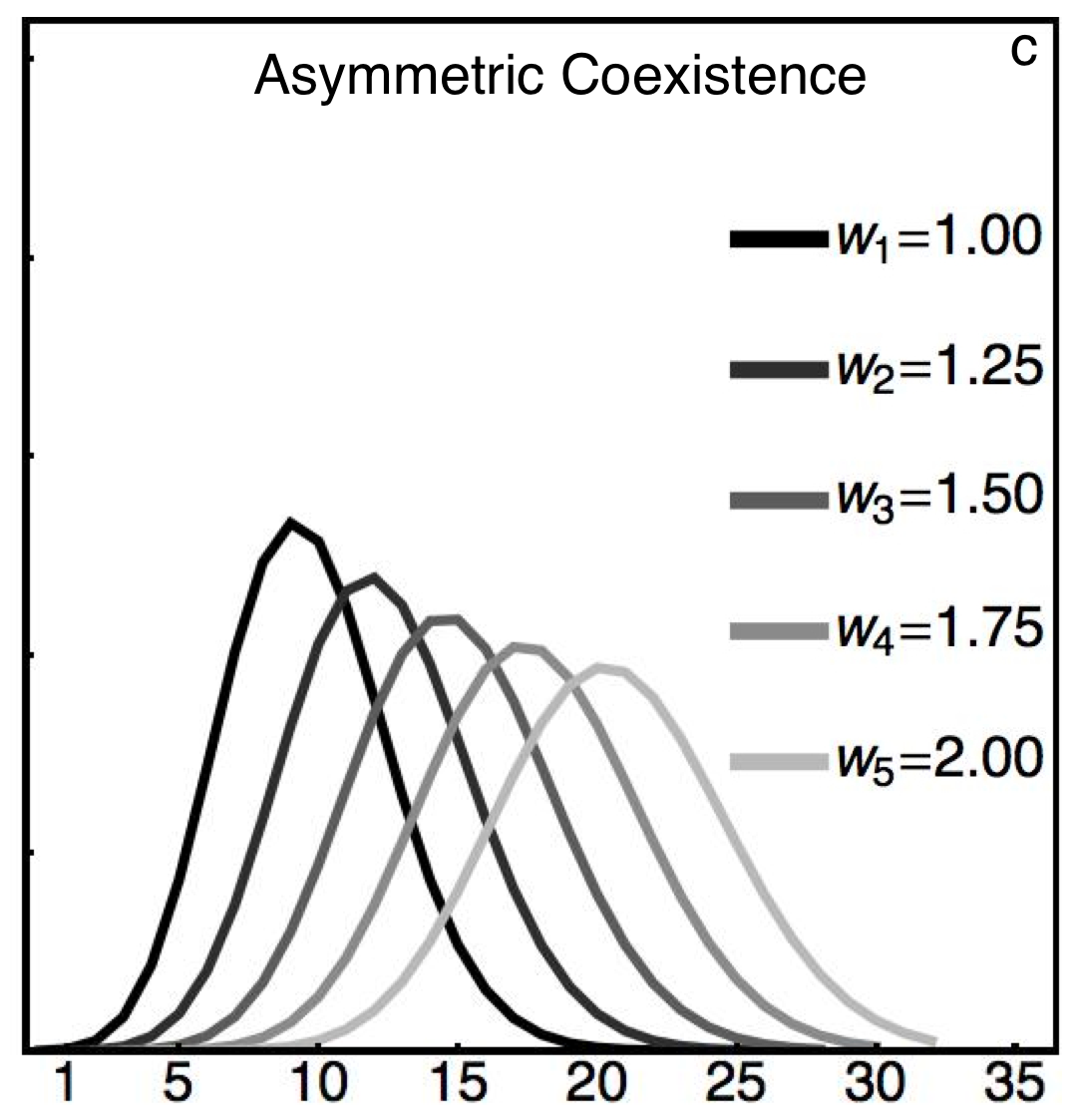}
}
\vskip0.0cm
\centerline{
\hskip0.03cm
\includegraphics[width=.595\textwidth]{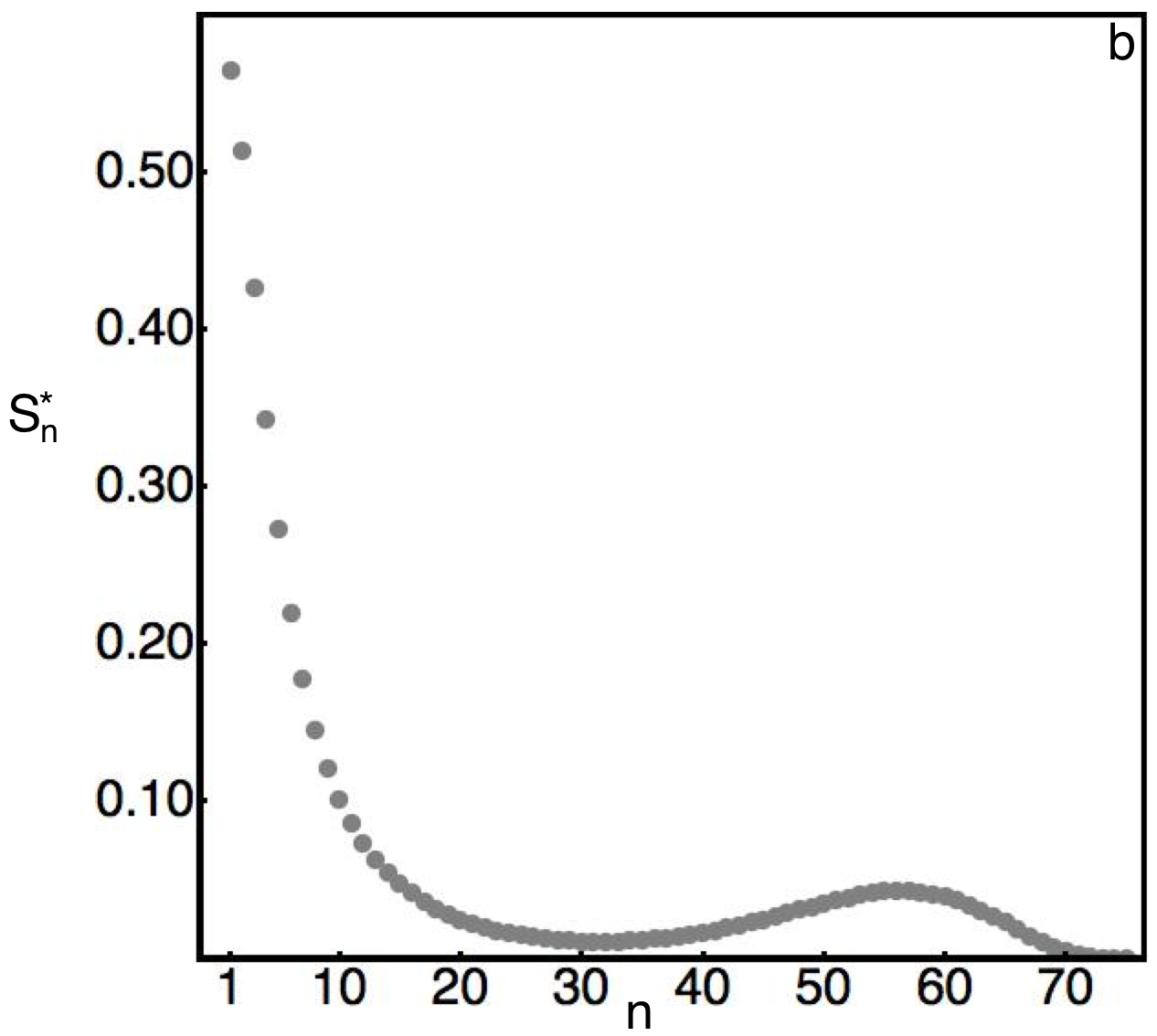}
\hskip-0.13cm
\includegraphics[width=.5000\textwidth]{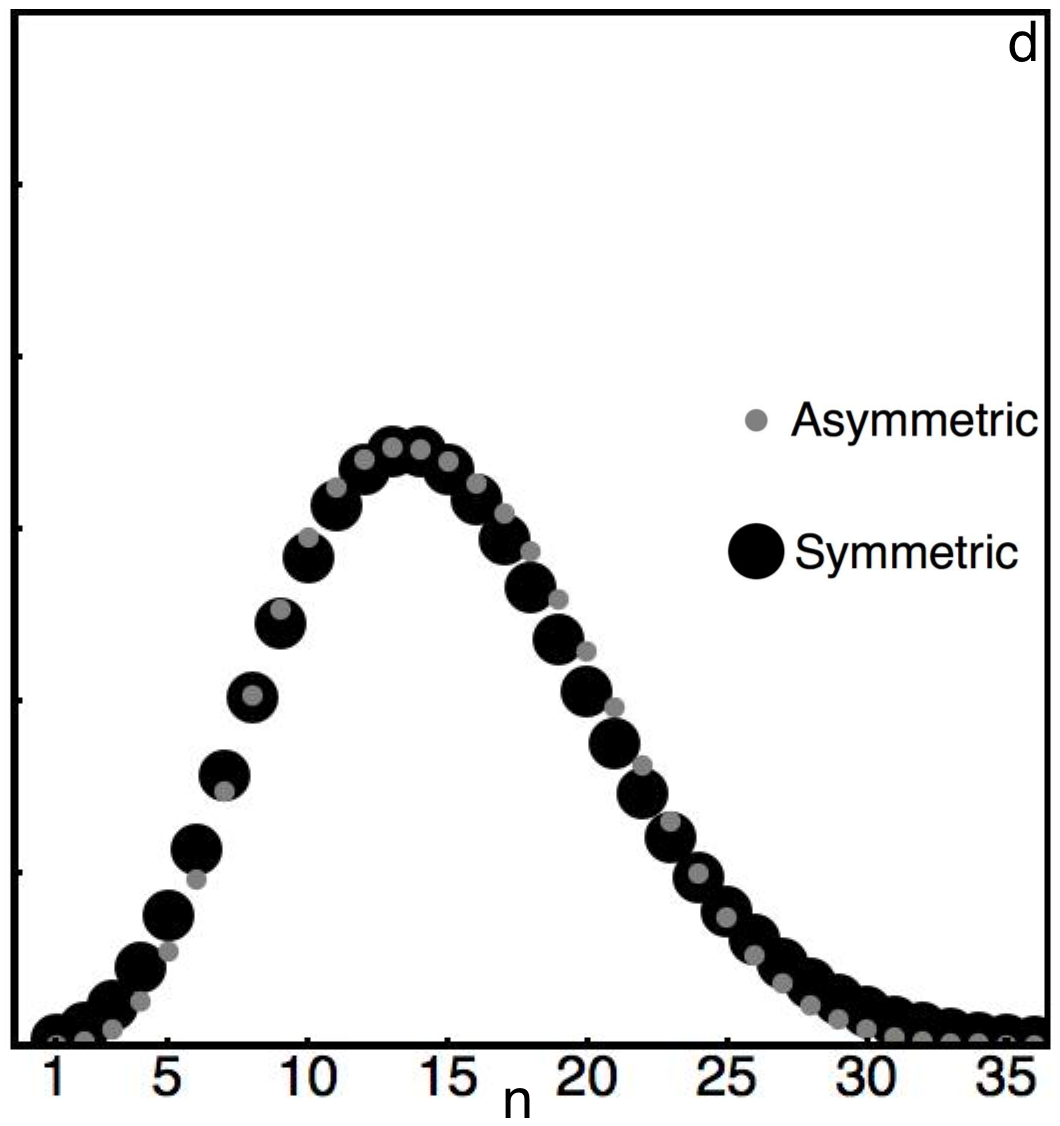}
}
\caption{Equilibrium abundance probabilities and corresponding SADs for an asymmetric community of $J_L=75$ individuals and $S=5$ species.  Non-neutral dynamics generate local deviations from the relative metacommunity abundance, $x_i=0.20$, common to all species.  (a) Dominance by the species with highest ecological fitness given competitive asymmetries in a community of uniform dispersal abilities (the $m_i=0.10$ for all $i$).  (b) The resulting bimodal SAD provides a strong indicator of non-neutral dynamics.  (c) Coexistence arising from mass-effects (the $m_i=0.90$ for all $i$).  (d) The resulting unimodal SAD closely resembles the SAD for a symmetric community of $J_L=75$ individuals and $S=5$ species where the $m_i=0.35$ and the $x_i=0.20$.}
\label{general}
\end{figure*}

Mass-effects allow for a soft breaking of the symmetry of neutral theory and provide a mechanism for multi-species coexistence.  In Fig.~\ref{general}, we present numerical results for the marginal equilibrium distributions of an asymmetric local community subsidized by a potentially neutral metacommunity, where the five species share a common relative abundance, $x_j$=0.2.  Although a single species may dominate due to a locally superior competitive ability (see Fig.~\ref{general}a), multi-species coexistence may arise, despite significant competitive asymmetries, due to high levels of immigration that tend to align local relative abundances with those in the metacommunity (see Fig.~\ref{general}c).  Despite the underlying asymmetric process, coexistence via mass-effects generates unimodal SADs that, given sampling errors in field data, may be indistinguishable from SADs due to neutral dynamics, as shown in Fig.~\ref{general}d.  This reinforces previous conclusions that the static, aggregate data in unimodal SADs cannot resolve the individual-level rules of engagement governing the origin and maintenance of biodiversity~\citep{Chave:2002p6357,Mouquet:2003p6380,Purves:2005,He:2005p7106,Chase:2005p6356,Walker:2007p4670,Doncaster:2009p6361}.  However, SADs with multiple modes are not uncommon in nature~\citep{Dornelas:2008p8337,Gray:2010p6748} and provide a strong indicator of non-neutral dynamics~\citep{Alonso:2008p4724}.  Fig.~\ref{general}b presents a bimodal SAD for an asymmetric local community with low levels of immigration.  

Each plot in Fig.~\ref{general} displays results for a relatively small community of $J_L=75$ individuals and $S=5$ possible species.  Sparse matrix methods were used to calculate the left eigenvector with zero eigenvalue for transition matrices of rank $\sim1.5{\rm M}$.  Obtaining stationary distributions for larger, more realistic communities poses a formidable numerical challenge.  This motivates a search for analytically tractable approximations to sampling distributions of the general theory.  

\section{An approximation to the sampling distribution}
\label{sc00}

The distribution $P^*_{\vec{n}}$ is stationary under Eq.~\ref{evolve1} if it satisfies the condition of detailed balance
\beq
T_{ij\vec{n}+\vec{e}_i-\vec{e}_j}P^*_{\vec{n}+\vec{e}_i-\vec{e}_j}\,=\,T_{ji\vec{n}}P^*_{\vec{n}},
\eeq{detailedbalance}
for all $i$ and $j$ such that $i\ne j$ and $\Theta_{ij}\ne 0$.  For general (g) large--$J_L$ communities where $S,w_k<<\sum_{l=1}^Sw_ln_l$ for all $k$, we will show that detailed balance is approximately satisfied by
\beq
P^{{\rm g}*}_{\vec{n}}\,=\,Z_{\rm g}^{-1}\binom{J_L}{n_1,\dots,n_S}\prod_{k=1}^Sw_k^{n_k}(1-m_k)^{n_k}\left(\phi_{k\vec{n}}x_k\right)_{n_k},
\eeq{gensampdistrib}
where 
\beqa
\phi_{k\vec{n}}\,=\,I_k\frac{(\sum_{l=1}^Sw_ln_l-w_k)/(J_L-1)}{\sum_{l=1}^Sw_lx_l},
\eeqa{phinvec}
where $(a)_n=\Gamma(a+n)/\Gamma(a)$ is the Pochhammer symbol, $Z_{\rm g}^{-1}$ is a normalization constant, and $I_k=m_k(J_L-1)/(1-m_k)$  is a generalization of the ``fundamental dispersal number"~\citep{Etienne:2005p3829}.  From the definition of $T_{ij\vec{n}}$ in Eq.~\ref{t}, we have
\beq
\frac{T_{ij\vec{n}+\vec{e}_i-\vec{e}_j}}{T_{ji\vec{n}}}\,=\,\frac{n_i+1}{n_j}\frac{w_j}{w_i}\frac{1-m_j}{1-m_i}\frac{n_j-1+
\phi_{j\vec{n}}x_j}{n_i+\phi_{i\vec{n}}x_i\left(1+\frac{w_i-w_j}{\sum_{l=1}^Sw_ln_l-w_i}\right)},
\eeq{tratioexact}
and assuming the form of $P^{{\rm g}*}_{\vec{n}}$ in Eq.~\ref{gensampdistrib}, we find
\beq
\frac{P^{{\rm g}*}_{\vec{n}}}{P^{{\rm g}*}_{\vec{n}+\vec{e}_i-\vec{e}_j}}\,=\,\frac{n_i+1}{n_j}\frac{w_j}{w_i}\frac{1-m_j}{1-m_i}\prod_{k=1}^S\frac{\left(\phi_{k\vec{n}}x_k\right)_{n_k}}{\left(\phi_{k\vec{n}}x_k\left(1+\frac{w_i-w_j}{\sum_{l=1}^Sw_ln_l-w_k}\right)\right)_{n_k+\delta_{ik}-\delta_{jk}}} \\.
\eeq{gensamptesta}
Now, for large--$J_L$ communities where $w_k<<\sum_{l=1}^Sw_ln_l$ for all $k$, the ratio $\epsilon_{ijk}\equiv (w_i-w_j)/(\sum_{l=1}^Sw_ln_l-w_k)$ is a small number.  Given $(a(1+\epsilon))_n\sim(a)_n+{\cal O}(\epsilon)$, we expand the right-hand-side of Eq.~\ref{gensamptesta} to obtain
\beq
\frac{P^{{\rm g}*}_{\vec{n}}}{P^{{\rm g}*}_{\vec{n}+\vec{e}_i-\vec{e}_j}}\,=\,\frac{T_{ij\vec{n}+\vec{e}_i-\vec{e}_j}}{T_{ji\vec{n}}}+\sum_{k=1}^S{\cal O}(\epsilon_{ijk}),
\eeq{gensamptestb}
which validates our assertion that Eq.~\ref{gensampdistrib} is an approximate sampling distribution of the general theory when $S<<\sum_{l=1}^Sw_ln_l$.  For communities of species that are symmetric (s) in ecological fitness but asymmetric in dispersal ability, Eq.~\ref{gensampdistrib} reduces to an exact sampling distribution 
\beq
P^{{\rm s}*}_{\vec{n}}\,=\,Z_{\rm s}^{-1}\binom{J_L}{n_1,\dots,n_S}\prod_{k=1}^S(1-m_k)^{n_k}\left(I_kx_k\right)_{n_k},
\eeq{gensampdistribfs}
that satisfies detailed balance without approximation.  Analogous distributions for general and fitness-symmetric metacommunities are provided in Appendix A.  However, in all of these results, the normalization constants must be calculated numerically.  This limits the utility of our sampling distributions in statistical analyses.  Can we find a non-neutral scenario that admits an approximate sampling distribution with an analytical expression for the normalization?



\section{Sampling nearly neutral communities}
\label{sc1}
As the species abundance vector evolves under Eq.~\ref{evolve1}, consider the dynamics of marginal abundance probabilities for a single focal species that deviates in ecological function from the surrounding, otherwise symmetric, community.  In particular, let the first element of $\vec{N}(\tau)$ be the marginal process, $N(\tau)$, over states $n\in(0,\dots,J_L)$, for the abundance of an asymmetric focal species with dispersal ability $m$, ecological fitness $w$, and relative metacommunity abundance $x$.  If all other species share a common dispersal ability $m_o$ and ecological fitness $w_o$, then the focal species gains an individual with probability
\beqa
g_{n}&\equiv&\sum_{i=2}^S T_{i1(n,n_2,\dots,n_S)} \nonumber \\
&=&\frac{J_L-n}{J_L}\left((1-m)\frac{w n}{w n+w_o(J_L-n-1)}+m\frac{w x}{wx+w_o(1-x)}\right), \nonumber \\
\eeqa{bn}
and loses an individual with probability
\beqa
r_{n} &\equiv& \sum_{i=2}^S T_{1i(n,n_2,\dots,n_S)} \nonumber \\
&=&\frac{n}{J_L}\left((1-m_o)\frac{w_o (J_L-n)}{w (n-1) + w_o(J_L-n)}+m_o\frac{w_o(1-x)}{w x+w_o(1-x)}\right), \nonumber \\
\eeqa{dn}
where we have used $\sum_{k=1}^Sx_k=1$.  These marginal transition probabilities do not depend separately on $w$ and $w_o$, but only on their ratio.  Without loss of generality, we redefine $w\equiv w/w_o$ to be the focal species' local {\it advantage} in ecological fitness.  Eqs.~\ref{bn} and \ref{dn}, which are independent of the abundances $(n_2,\dots,n_S)$, suggest a univariate birth-death process for the marginal dynamics of the asymmetric species governed by the master equation 
\beqa
\frac{dP_n}{d\tau}&=&g_{n-1}\Theta(n-1)P_{n-1}+r_{n+1}\Theta(J_L-(n+1))P_{n+1} \nonumber \\
&&-(g_n\Theta(J_L-(n+1))+r_n\Theta(n-1))P_n,
\eeqa{margprocessasymm}
and we formally derive this result from Eq.~\ref{evolve1} in Appendix B.
Given the well-known stationary distribution of Eq.~\ref{margprocessasymm}
\beq
P_n^* \,=\, P_0^* \prod_{i=0}^{n-1}\frac{g_{i}}{r_{i+1}}.
\eeq{Pn}
we find an exact result for the stationary abundance probabilities of the focal species in a nearly neutral (nn) community
\beq
P_n^{{\rm nn}*}\,=\,Z \binom{J_L}{n}\eta^n \frac{{\rm B}(\lambda+n,\xi-n)}{{\rm B}(\lambda,\xi)},
\eeq{PnNN}
where ${\rm B}(a,b)=\Gamma(a)\Gamma(b)/\Gamma(a+b)$ is the beta-function
\beq
Z^{-1}\,=\, {}_2F_1(-J_L,\lambda;1-\xi;\eta),
\eeq{NPL}
and
\beqa
\eta&=&w\frac{1-m+x(w-1)}{1-m_o w+x(w-1)}, \nonumber \\
\lambda&=&\frac{(J_L-1)mx}{1-m+x(w-1)}, \nonumber \\
\xi&=& 1+\frac{(J_L-1)(1-x w m_o+x(w-1))}{1  - w m_o + x(w-1)}.
\eeqa{abcgen}
For the asymmetric focal species, this is an exact result of the general model, Eq.~\ref{evolve1}, that holds for nearly neutral local communites with any number of additional species.  Eq.~\ref{PnNN} may be classified broadly as a generalized hypergeometric distribution or more specifically as an exponentially weighted P\'olya distribution~\citep{Kemp:1968tSUDDT,Johnson:1992tUDD}. 

In the absence of dispersal limitation, Eq.~\ref{PnNN} becomes
\beq
\lim_{m,m_o\rightarrow1}P_n^{{\rm nn}*}\,=\,\left(\frac{1}{1+x(w-1)}\right)^{J_L}\binom{J_L}{n}(wx)^n (1-x)^{J_L-n},
\eeq{PnNNlimit}
where the identity ${\rm B}(a,b){\rm B}(a+b,1-b)=\pi/(a\sin(\pi b))$ has been used.  This is a weighted binomial distribution with expected abundance $wxJ_L/(1+x(w-1))$ and variance $wx(1-x)J_L/(1+x(w-1))^2$.  In the neutral, or symmetric, limit where $w=1$, Eq.~\ref{PnNNlimit} reduces to a binomial sampling of the metacommunity, {\it sensu}~\citet{Etienne:2005p3829}.  

In the presence of dispersal limitation, we evaluate $\Sigma_{n=1}^{J_L} n P_n^{{\rm nn}*}$ to obtain the expected abundance
\beqa
{\rm E}[N^*]&=&\eta\frac{\partial}{\partial\eta}\log Z \nonumber \\
&=&\frac{J_L\lambda\eta}{\xi-1}
\frac{
{}_2F_1\left(1-J_L,1+\lambda;2-\xi;\eta\right)}
{
{}_2F_1\left(-J_L,\lambda;1-\xi;\eta\right)},
\eeqa{avgn}
where $N^*\equiv\lim_{\tau\to\infty}N(\tau)$.  The variance of the stationary distribution is given by
\beqa
{\rm Var}[N^*]&=& \eta \frac{\partial}{\partial\eta} \eta \frac{\partial}{\partial\eta} \log Z \nonumber \\
&=&{\rm E}[N^{*2}]-{\rm E}[N^*]^2,
\eeqa{Var}
and we evaluate $\Sigma_{n=1}^{J_L} n^2 P_n^{{\rm nn}*}$ to obtain
\beqa
{\rm E}[N^{*2}]\,=\,\frac{J_L\lambda\eta}{\xi-1}
\frac{
{}_3F_2\left(1-J_L,1+\lambda,2;2-\xi,1;\eta\right)}
{
{}_2F_1\left(-J_L,\lambda;1-\xi;\eta\right)}. 
\eeqa{avgn2}
In Eqs.~\ref{avgn} and \ref{Var}, the normalization of Eq.~\ref{NPL} generates central moments for the abundance distribution and plays a role analogous to the grand partition function of statistical physics.  Recent studies have demonstrated the utility of partition functions in extensions of Hubbell's neutral theory~\citep{ODwyer:2009p4394, ODwyer:2009p6371}.

\begin{figure*}
\centerline{
\includegraphics[width=.58625\textwidth]{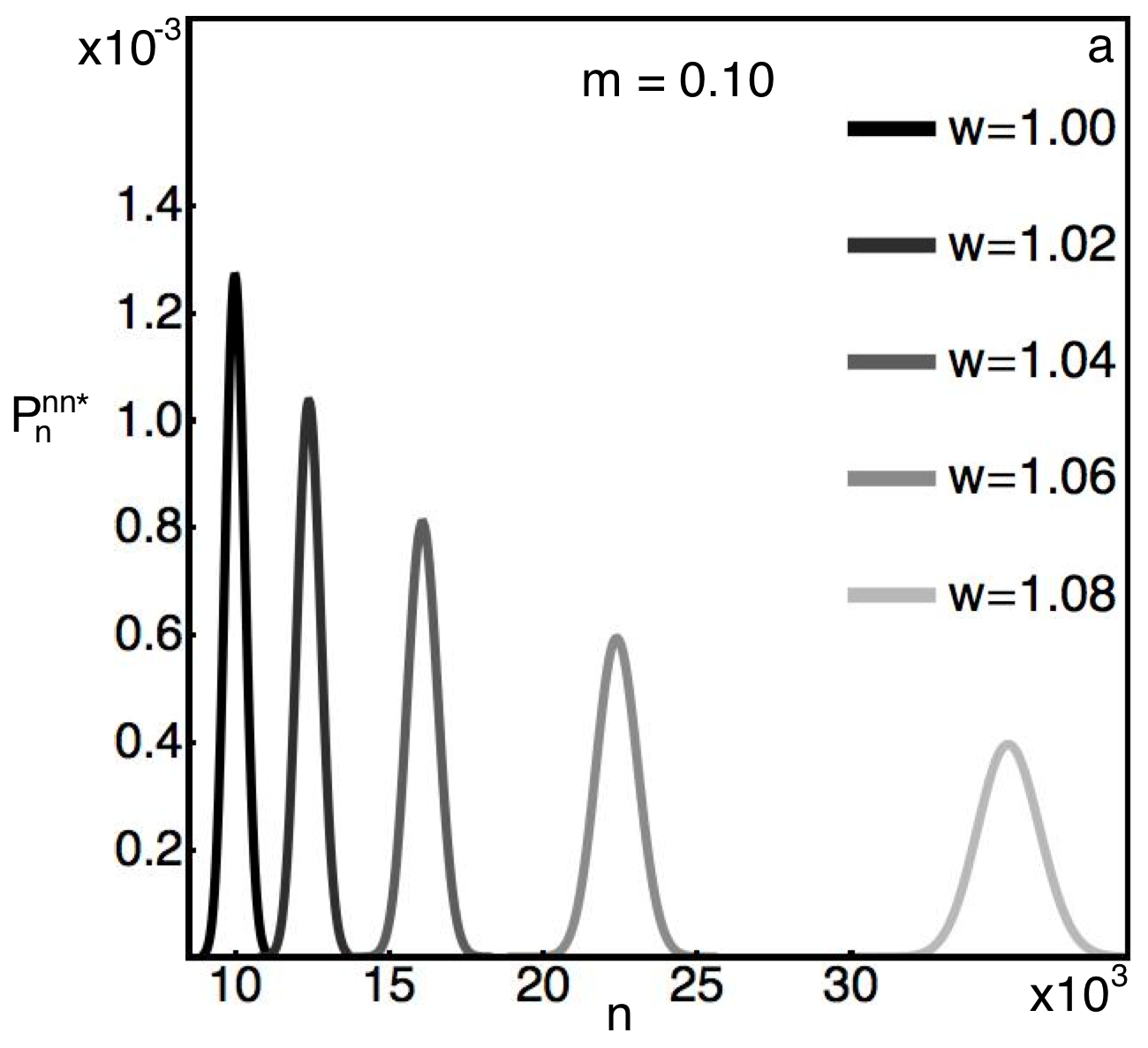}
\hskip-0.15cm
\includegraphics[width=.5\textwidth]{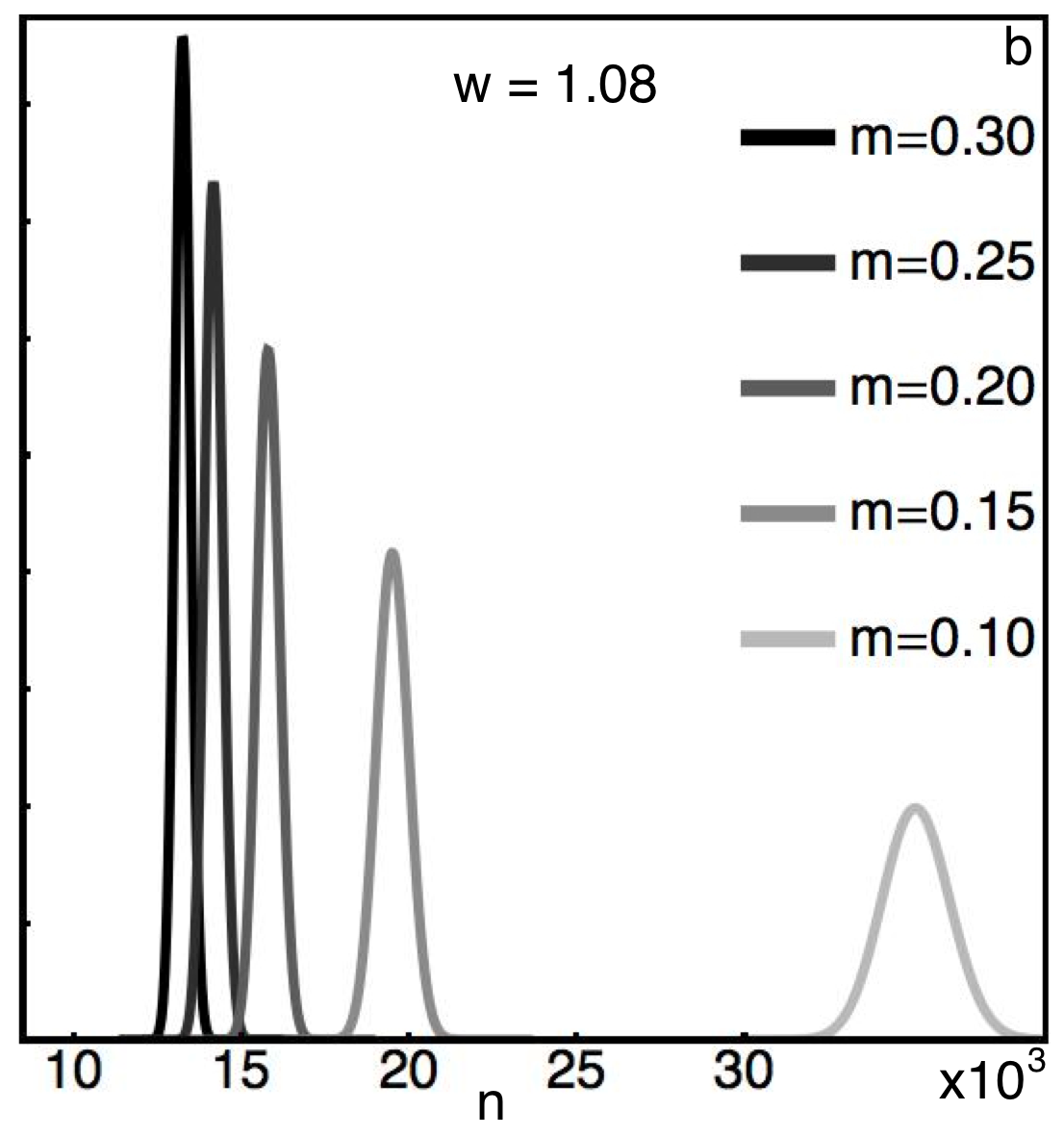}
}
\caption{Novel asymptotic expansions of hypergeometric functions have been used to plot marginal equilibrium abundance probabilities for the asymmetric focal species, with relative metacommunity abundance $x=0.01$, in a nearly neutral local community of $J_L$=1M individuals.  (a) Dominance with rising advantage in ecological fitness, as indicated for each curve.  Here, all species are symmetric in their dispersal ability ($m=m_o=0.10$).  (b) Dispersal mitigates the advantage in ecological fitness ($w=1.08$) of the asymmetric focal species.  All other species share a common dispersal ability of $m_o=0.10$.}
\label{asymp}
\end{figure*}

For large--$J_L$ communities, evaluation of the hypergeometric functions in Eqs.~\ref{PnNN}, \ref{avgn}, and \ref{avgn2} is computationally expensive.  To remove this barrier, one of us (N.M.T.) has derived novel asymptotic expansions (see Appendix C).  We use these expansions to plot the stationary abundance probabilities for $J_L=1$M.  In Fig.~\ref{asymp}a, small local advantages in ecological fitness generate substantial increases in expected abundance over the neutral prediction.  Hubbell found evidence for these discrepancies in Manu forest data and referred to them as ``ecological dominance deviations"~\citep{PHubbell:2001p4284}.  Hubbell also anticipated that dispersal effects would mitigate advantages in ecological fitness~\citep{PHubbell:2001p4284}. The right panel of Fig.~\ref{asymp} demonstrates, once again, that enhanced mass-effects due to increased dispersal ability may inhibit the dominance of a locally superior competitor by compelling relative local abundance to align with relative metacommunity abundance.

An approximation to the multivariate sampling distribution of nearly neutral local communities is constructed in Appendix B
\beq
P^{{\rm nn}*}_{\vec{n}}\,=\,Z\binom{J_L}{n,n_2,\dots,n_S}\eta^{n}\frac{{\rm B}(\lambda+n,\xi-n)}{{\rm B}(\lambda,\xi)}\frac{1}{((1-x)\phi_{on})_{J_L-n}}\prod_{i=2}^S(\phi_{on} x_i)_{n_i}, 
\eeq{PnvecNN}
where
\beq
\phi_{on}\,=\,I_o\frac{1+n(w-1)/(J_L-1)}{1+x(w-1)}.
\eeq{defphi}
A related approximation for the sampling distribution of nearly neutral metacommunities is derived in Appendix A.  In the absence of dispersal limitation, Eq.~\ref{PnvecNN} becomes
\beq
\lim_{m,m_o\rightarrow1}P^{{\rm nn}*}_{\vec{n}}\,=\,\left(\frac{1}{1+x(w-1)}\right)^{J_L}\binom{J_L}{n,n_2,\dots,n_S}(wx)^n\prod_{i=2}^Sx_i^{n_i},
\eeq{PnvecNNlimit}
where we have used $(a)_n\sim a^n+{\cal O}(a^{n-1})$ for large $a$.  Finally, in the symmetric limit, Eq.~\ref{PnvecNNlimit} reduces to a simple multinomial sampling of the metacommunity, as expected.

\begin{figure*}
\centerline{
\includegraphics[width=0.8\textwidth]{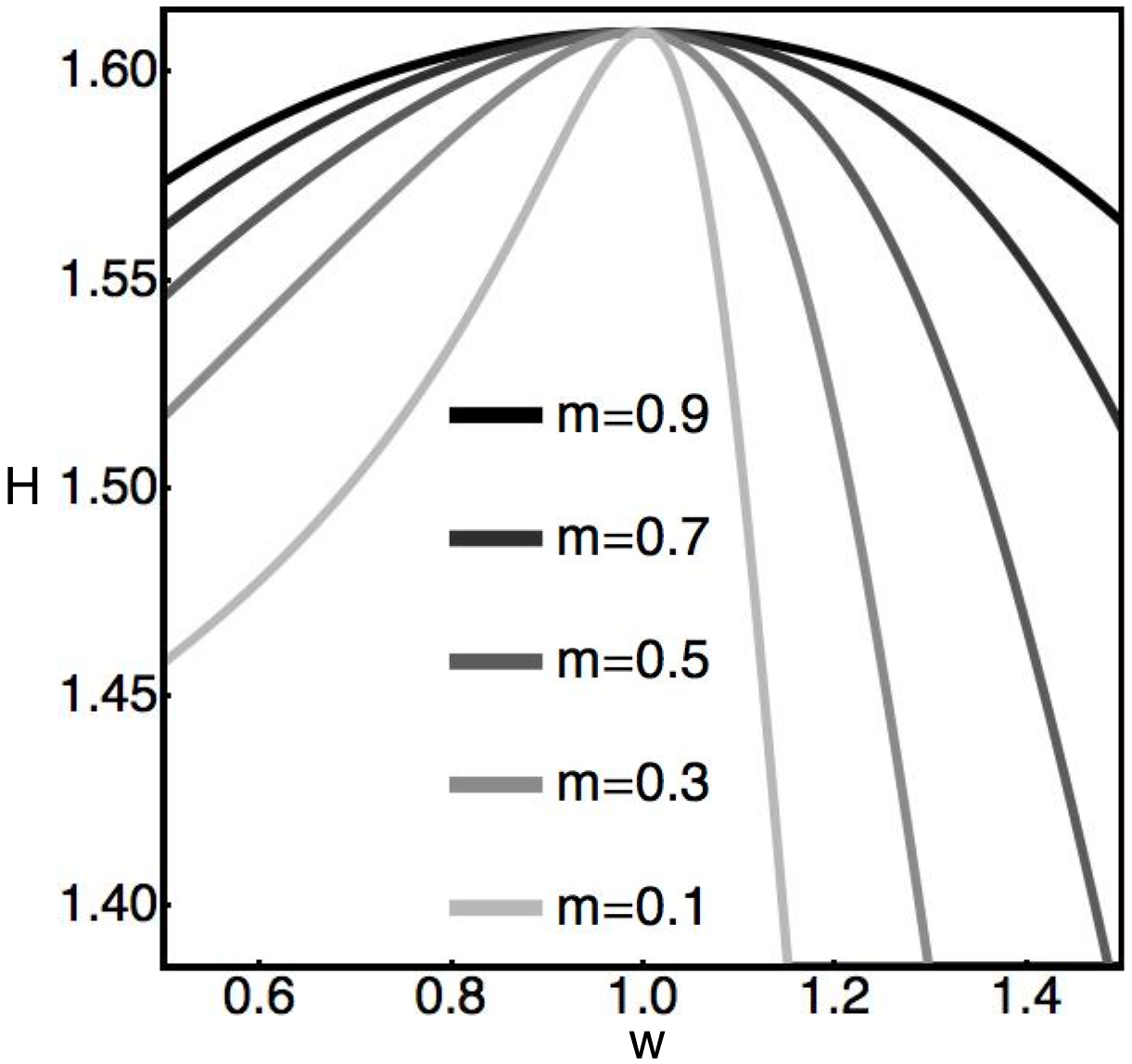}
}
\caption{Plots of the information-theoretic diversity metric, $H$, for a nearly neutral local community of $J_L=75$ individuals and $S=5$ species over various values of the ecological fitness advantage, $w$, and dispersal ability, $m$.  The symmetric species share a common dispersal ability of $m_o=0.1$, and all species share a common relative metacommunity abundance such that $x=x_i=0.2$.  Diversity peaks where expected local abundances are equivalent, and this occurs in the symmetric limit, given by $w=1$ and $m=m_o$.  The asymmetries in dispersal ability shown here have a negligible impact on diversity when all species are symmetric in ecological fitness, so each curve in Fig.~\ref{div} peaks at approximately the same value of $H$ near $w=1$.  Away from the peak, declines in diversity are regulated by mass-effects, with more gradual declines at higher values of $m$.}
\label{div}
\end{figure*}

To illustrate the impacts of an asymmetric species on the diversity of an otherwise symmetric local community, Fig.~\ref{div} plots Shannon's Index of diversity 
\beq
H=-\frac{{\rm E}[N^*]}{J_L}\log \frac{{\rm E}[N^*]}{J_L}-\sum_{i=2}^S\frac{{\rm E}[N_i^*]}{J_L}\log \frac{{\rm E}[N_i^*]}{J_L},
\eeq{Hdiv} 
for various values of the ecological fitness advantage, $w$, and dispersal ability, $m$, in a nearly neutral community of $S=5$ species and $J_L=75$ individuals.  All five species share a common relative metacommunity abundance, $x=x_i=0.2$, so given the exact result for ${\rm E}[N^*]$ in Eq.~\ref{avgn}, we know immediately that ${\rm E}[N_i^*]=(J_L-{\rm E}[N^*])/(S-1)$ for the remaining symmetric species.  Note that $H$ is maximized where all abundances are equivalent, such that ${\rm E}[N^*]/J_L={\rm E}[N_i^*]/J_L=x_i$.  As can be seen from the next section, this relation holds in the neutral limit where $w=1$ and $m=m_o=0.1$, but small asymmetries in dispersal ability have a negligible impact on diversity when all species are symmetric in ecological fitness.  Therefore, each curve in Fig.~\ref{div} peaks near $w=1$ at approximately the same value of $H$.  Away from $w=1$, the declines in diversity are regulated by mass-effects, with more gradual declines at higher values of $m$.  

\section{Recovering the sampling distribution of neutral theory}
\label{sc2}
In a perfectly symmetric local community, the stochastic dynamics for each species differ solely due to variations in relative metacommunity abundances, the $x_i$.  In particular, if $m_j=m$ and $w_j=1$ for all $j$ in Eq.~\ref{t}, we recover the multivariate transition probabilities for a neutral sampling theory of local communities, as suggested on p.~287 of Hubbell's book~\citep{PHubbell:2001p4284}.  Similarly, in the symmetric limit of Eq.~\ref{margprocessasymm} where $m=m_o$ and $w=1$, we recover the marginal dynamics for neutral (n) local communities with stationary distribution~\citep{McKane:2000p4318}
\beq
P^{{\rm n}*}_n\,=\,\binom{J_L}{n}\frac{{\rm B}(I x+n,J_L+I(1-x)-n)}{{\rm B}(I x,I(1-x))},
\eeq{PnNeutral}
where $I=m(J_L-1)/(1-m)$.  This result follows from the symmetric limit of Eq.~\ref{PnNN} after applying the identity $\Gamma(a)\Gamma(1-a)=\pi/\sin(\pi a)$.  The expected abundance and variance are obtained from the symmetric limits of Eqs.~\ref{avgn} and \ref{Var}, respectively, after applying the identities in Eqs.~C.1.0.1 and C.2.2.2 
\begin{eqnarray}
{\rm E}[N^*]&=& xJ_L, \label{avgnNeutral} \\
{\rm Var}[N^*]&=&x(1-x)J_L \frac{J_L+I}{1+I}. \label{VarNeutral}
\end{eqnarray}
Finally, the symmetric limits of Eqs.~\ref{gensampdistrib}, \ref{gensampdistribfs}, and \ref{PnvecNN} all yield the stationary sampling distribution for a neutral local community~\citep{Etienne:2005p3829,Etienne:2007p5640}
\beq
P^{{\rm n}*}_{\vec{n}}\,=\,\binom{J_L}{n_1,\dots,n_S}\frac{1}{(I)_{J_L}}\prod_{i=1}^S(Ix_i)_{n_i}.
\eeq{PnecNeutral}
In the special case of complete neutrality, Eq.~\ref{PnecNeutral} is an exact result of the general model, Eq.~\ref{evolve1}.  This sampling distribution continues to hold when the assumptions of zero-sum dynamics and stationarity are relaxed~\citep{Etienne:2007p5640,Haegeman:2008p6364}.

\section{Discussion}
\label{sc3}
We have developed a general sampling theory that extends Hubbell's neutral theory of local communities and metacommunities to include asymmetries in ecological fitness and dispersal ability.  We anticipate that a parameterization of additional biological complexity, such as asymmetries in survivorship probabilities or differences between the establishment probabilities of local reproduction and immigration, may be incorporated without significant changes to the structure of our analytical results.  Although the machinery is significantly more complicated for asymmetric theories than their symmetric counterparts, some analytical calculations remain tractable.  We find approximate sampling distributions for general and nearly neutral communities that yield Hubbell's theory in the symmetric limit.  Our fully normalized approximation in the nearly neutral case may provide a valuable statistical tool for determining the degree to which an observed SAD is consistent with the assumption of complete neutrality.  To facilitate a Bayesian analysis, we have enabled rapid computation of the required hypergeometric functions by deriving previously unknown asymptotic expansions.

\section*{Acknowledgments}
\label{ack}
We gratefully acknowledge the insights of two anonymous reviewers.  This work is partially supported by the James S. McDonnell Foundation through their Studying Complex Systems grant (220020138) to W.F.F.  N.M.T. acknowledges financial support from Gobierno of Navarra, Res.~$07/05/2008$ and Ministerio de Ciencia e Innovaci\'on, project MTM2009-11686.


\appendix

\makeatletter
\@addtoreset{equation}{subsubsection}
\makeatother
\renewcommand{\theequation}{A.\arabic{equation}}

\section{Sampling asymmetric metacommunities}
\label{othapp}

The analytical insights of \citet{Etienne:2007p5640} suggest a clear prescription for translating local community dynamics into metacommunities dynamics in the context of Hubbell's unified neutral theory of biodiversity and biogeography~\citep{PHubbell:2001p4284}:  replace probabilities of immigration, $m_j$, with probabilities of speciation, $\nu_j$; assume $x_j\sim 1/S_T+{\cal O}(1/S_T^2)$ for all $j$, where $S_T$ is the total number of species that could possibly appear through speciation events; and consider asymptotics as $S_T$ becomes large.

Following this recipe, we translate the transition probabilities for asymmetric local communities, Eq.~\ref{t}, into the transition probabilities for asymmetric metacommunities ($M$)
\beq
T^M_{ij\vec{n}} \,=\, \frac{n_i}{J_M}\left((1-\nu_j)\frac{w_jn_j}{\sum_{k=1}^{S_T}w_kn_k-w_i}+\nu_j\frac{w_j}{\sum_{k=1}^{S_T}w_k}+{\cal O}\left(\frac{1}{S_T}\right)\right),
\eeq{tM}
where $J_M$ is the number of individuals in the metacommunity, $w_j/\sum_{k=1}^{S_T}w_k$ is the probability that an individual of species $j$ establishes following a speciation event, and
\beq
\Theta_{ij}^M\,=\,\Theta(J_M-(n_i+1))\Theta(n_j-1).
\eeq{ThetaM}
Metacommunity dynamics are governed by the master equation
\beq
\frac{dP^M_{\vec{n}}}{d\tau}\,=\,\sum_{i=1}^{S_T}\sum_{j=1,j\ne i}^{S_T}\left(T^M_{ij\vec{n}+\vec{e}_i-\vec{e}_j}P^M_{\vec{n}+\vec{e}_i-\vec{e}_j}-T^M_{ji\vec{n}}P^M_{\vec{n}} \right) \Theta_{ij}^M.
\eeq{evolve1M}
If $\nu_j>0$ for all $j$, there are no absorbing states, so for large--$S_T$, there is a nonzero probability that any given species $j$ exists.  Analogous develops to those in Section~\ref{sc00} show that detailed balance in the general theory is approximated by
\beq
P^{{\rm g,M}*}_{\vec{n}}\,=\,Z_{\rm g,M}^{-1}\binom{J_M}{n_1,\dots,n_{S_T}}\prod_{k=1}^{S_T}w_k^{n_k}(1-\nu_k)^{n_k}\left(\phi^{\rm M}_{k\vec{n}}\right)_{n_k},
\eeq{metagensampdistrib}
where 
\beqa
\phi^{\rm M}_{k\vec{n}}\,=\,\theta_k\frac{(\sum_{l=1}^{S_T}w_ln_l-w_k)/(J_M-1)}{\sum_{l=1}^{S_T}w_l},
\eeqa{metaphinvec}
and $\theta_k=\nu_k(J_M-1)/(1-\nu_k)$ is the generalization of Hubbell's ``fundamental biodiversity number"~\citep{PHubbell:2001p4284}.  The fitness-symmetric (s) distribution
\beq
P^{{\rm s,M}*}_{\vec{n}}\,=\,Z_{\rm s,M}^{-1}\binom{J_M}{n_1,\dots,n_{S_T}}\prod_{k=1}^{S_T}(1-\nu_k)^{n_k}\left(\theta_k/S_T\right)_{n_k},
\eeq{metagensampdistribfs}
satisfies detailed balance up to ${\cal O}(1/S_T)$.

For the special case of nearly neutral metacommunities, we translate the marginal dynamics for an asymmetric species in an otherwise symmetric local community into the marginal dynamics for an asymmetric species in an otherwise symmetric metacommunity.  The transition probabilities are 
\beqa
g^M_{n}&=&\frac{J_M-n}{J_M}\left((1-\nu)\frac{w n}{J_M+n(w-1)-1}+\nu\frac{w}{S_T}+{\cal O}\left(\frac{1}{S_T}\right)\right), \nonumber \\
r^M_{n} &=&\frac{n}{J_M}\left((1-\nu_o)\frac{J_M-n}{J_M+n(w-1)-w}+\nu_o\left(1-\frac{w}{S_T}\right)+{\cal O}\left(\frac{1}{S_T}\right)\right), \nonumber \\
\eeqa{bndnM}
where the asymmetric focal species has speciation probability $\nu$ and enjoys an ecological fitness advantage, $w$, over all other species, which share a common probability of speciation $\nu_o$.  If $n=0$ is an accessible state, then as $S_T$ becomes large and $w$ remains finite, the equilibrium probability of observing the asymmetric species approaches zero.  However, if we assume that the asymmetric species is identified and known to exist at nonzero abundance levels, the stationary distribution is 
\beq
P_n^{{\rm nn},M*}\,=\,Z_M \binom{J_M}{n}\eta_M^n \frac{{\rm B}(\lambda_M+n,\xi_M-n)}{{\rm B}(\lambda_M,\xi_M)},
\eeq{PnNNM}
with 
\beq
Z_M^{-1}\,=\, {}_2F_1(-J_M,\lambda_M;1-\xi_M;\eta_M)-1,
\eeq{NPLM}
and
\beqa
\eta_M&=&w\frac{(J_M-1)(J_M+\theta_o-1)}{(J_M-\theta_o(w-1)-1)(J_M+\theta-1)}+{\cal O}\left(\frac{1}{S_T}\right), \nonumber \\
\lambda_M&=&\frac{\theta}{S_T}+{\cal O}\left(\frac{1}{S^2_T}\right), \nonumber \\
\xi_M&=& 1+\frac{(J_M-1)(J_M+\theta_o-1)}{J_M-\theta_o(w-1)-1}+{\cal O}\left(\frac{1}{S_T}\right).
\eeqa{abcgenM}
where $\theta=(J_M-1)\nu/(1-\nu)$ and $\theta_o=(J_M-1)\nu_o/(1-\nu_o)$ are Hubbell's ``fundamental biodiversity numbers" for the asymmetric species and all other species, respectively.

An approximate multivariate stationary distribution is obtained in an identical manner to the derivation of Eq.~\ref{PnvecNN}
\beqa
P^{{\rm nn},M*}_{\vec{n}}&=&Z_M\binom{J_M}{n,n_2,\dots,n_{S_T}}\eta_M^n\frac{{\rm B}(\lambda_M+n,\xi_M-n)}{{\rm B}(\lambda_M,\xi_M)} \nonumber \\
&&\times\frac{1}{((1-1/S_T)\phi_{n,M})_{J_M-n}}\prod_{i=2}^{S_T}(\phi_{n,M}/S_T)_{n_i}, 
\eeqa{PnvecNNM}
where 
\beq
\phi_{n,M}\,=\,\theta_o\left(1+\frac{n(w-1)}{J_M-1}\right)+{\cal O}\left(\frac{1}{S_T}\right).
\eeq{phinM}

We now propose a modest extension to the prescription in \citet{Etienne:2007p5640} for converting multivariate distributions over labelled abundance vectors to distributions over unlabelled abundance vectors.  Because the asymmetric focal species has been identified and is known to exist with abundance $n>0$, this species must be labelled, while all other species are equivalent and may be unlabelled.  Therefore, we aim to transform Eq.~\ref{PnvecNNM} into a multivariate distribution over the ``mostly unlabelled" states $\hat{\vec{n}}=(n,\hat{n}_2,\dots,\hat{n}_S)$, where $S$ is the number of species observed in a sample and each $(\hat{n}_2,\dots,\hat{n}_S)$ is an integer partition of $J_M-n$.  (To provide an example, if $J_M=3$, four distinct states are accessible:  $(3)$ with $S=1$, $(2,1)$ with $S=2$, $(1,2)$ with $S=2$, and $(1,1,1)$ with $S=3$.)  The conversion is given by
\beq
P^{{\rm nn},M*}_{\hat{\vec{n}}}\,=\,\frac{(S_T-1)!}{\prod_{i=0}^{J_M-n}\hat{\Phi}_i!}P^{{\rm nn},M*}_{\vec{n}},
\eeq{PnhatvecM}
where $\hat{\Phi}_i$ is the number of elements in $(\hat{n}_2,\dots,\hat{n}_S)$ equal to $i$.  Note that $\hat{\Phi}_0=S_T-1-(S-1)$.  Taking the leading behavior for large--$S_T$, we obtain a modification of the~\citet{Ewens:1972p4815} sampling distribution appropriate to nearly neutral metacommunities
\beqa
\hskip-0.3cm\lim_{S_T\to\infty}P^{{\rm nn},M*}_{\hat{\vec{n}}}\hskip-0.3cm&=&\hskip-0.3cm\lim_{S_T\to\infty}\frac{(S_T-1)!}{(S_T-S)!}\frac{J_M!}{n\prod_{i=2}^S\hat{n}_i\prod_{i=1}^{J_M-n}\hat{\Phi}_i!} \nonumber \\
&&\times\frac{\eta_M^n(\theta/S_T)_n(\xi_M)_{-n}}{(n-1)!({}_2F_1(-J_M,\theta/S_T;1-\xi_M;\eta_M)-1)}\nonumber \\
&&\times\frac{1}{((1-1/S_T)\phi_{n,M})_{J_M-n}}\prod_{i=2}^S\frac{(\phi_{n,M}/S_T)_{\hat{n}_i}}{(\hat{n}_i-1)!} \nonumber \\ 
\nonumber \\
&=&\hskip-0.3cm\lim_{S_T\to\infty}\frac{(S_T-1)!}{(S_T-S)!}\left(\frac{1}{S_T}\right)^{S-1}\frac{J_M!}{n\prod_{i=2}^S\hat{n}_i\prod_{i=1}^{J_M-n}\hat{\Phi}_i!} \nonumber \\
&&\times\frac{\theta/S_T}{{}_2F_1(-J_M,\theta/S_T;1-\xi_M;\eta_M)-1}\eta_M^n(\xi_M)_{-n}\frac{\phi_{n,M}^{S-1}}{(\phi_{n,M})_{J_M-n}}\nonumber \\
\nonumber \\
&=&\hskip-0.3cm\hat{Z}_M\frac{J_M!}{n(J_M-n)!}\eta_M^n(\xi_M)_{-n}\frac{(J_M-n)!}{\prod_{i=2}^S\hat{n}_i\prod_{i=1}^{J_M-n}\hat{\Phi}_i!}\frac{\phi_{n,M}^{S-1}}{(\phi_{n,M})_{J_M-n}}, \nonumber \\
\eeqa{PnhatvecMlimit}
where $(a)_0=1$ allows us to take a product over the observed species, $S$, rather than the total number of possible species, $S_T$, in the first expression; $(z)_n/(n-1)!\sim z+{\cal O}(z^2)$ as $z$ approaches 0 for $n>0$ has been used to obtain the second expression; l'H\^opital's rule along with $\lim_{b\to 0}\partial {}_2F_1(a,b;c;z)/\partial b=az{}_3F_2(a+1,1,1;c+1,2;z)/c$ has been used to obtain the third expression; and
\beq
\hat{Z}_M^{-1}\,=\,\frac{J_M\eta_M}{\xi_M-1} {}_3F_2(1-J_M,1,1;2-\xi_M,2;\eta_M),
\eeq{NpPLM}
with asymptotics of the hypergeometric function provided in \S\ref{F21forF32} and \S\ref{F32case}.  In the neutral limit, we obtain a modification to the Ewens sampling distribution for the scenario where a single species is labelled and guaranteed to exist
\beq
\lim_{S_T\to\infty}P^{{\rm n},M*}_{\hat{\vec{n}}}\,=\,\left(\sum_{i=1}^{J_M}\frac{\theta}{\theta+i}+\frac{J_M}{\theta+J_M}\right)^{-1}\frac{J_M!}{n\prod_{i=2}^S\hat{n}_i\prod_{i=1}^{J_M-n}\hat{\Phi}_i!}\frac{\theta^S}{(\theta)_{J_M}}.
\eeq{PnhatvecMlimitNeutral}
Converting this result to a distribution over the ``fully unlabelled" states $\hat{\hat{\vec{n}}}=\left(\hat{\hat{n}}_1,\dots,\hat{\hat{n}}_S\right)$, we multiply by 
\beq
 \left(\sum_{i=1}^{J_M}\frac{\theta}{\theta+i}+\frac{J_M}{\theta+J_M}\right)\frac{\prod_{i=1}^{J_M-n}\hat{\Phi}_i!}{\prod_{i=1}^{J_M}\hat{\hat{\Phi}}_i!},
\eeq{convert}
and recover the Ewens sampling distribution~\citep{Ewens:1972p4815}, which is also the sampling distribution for Hubbell's metacommunity theory~\citep{PHubbell:2001p4284}.


\makeatletter
\@addtoreset{equation}{subsubsection}
\makeatother
\renewcommand{\theequation}{B.\arabic{equation}}

\section{Marginal dynamics for the local community}
\label{app}

We first demonstrate that the marginal dynamics of the asymmetric species in Eq.~\ref{margprocessasymm} can be derived from the multivariate dynamics of Eq.~\ref{evolve1}.  Let 
\beq
\sum^{J_L-n}\delta\,\equiv\,\sum_{n_2=0}^{J_L-n} \dots \sum_{n_S=0}^{J_L-n}\delta(J_L-n-n_2-\dots-n_S),
\eeq{defsumdeltaasymm}
so that the marginal distribution for the asymmetric species is given by
\beq
P_n\,=\,\sum^{J_L-n}\delta P_{\vec{n}}.
\eeq{pnexampleasymm}
Applying Eq.~\ref{defsumdeltaasymm} to both sides of Eq.~\ref{evolve1}, we obtain
\beqa
\frac{dP_n}{d\tau}&=&\sum^{J_L-n}\delta \sum_{i=1}^S\sum_{j=1,j\ne i}^S\left(T_{ij\vec{n}+\vec{e}_i-\vec{e}_j}P_{\vec{n}+\vec{e}_i-\vec{e}_j}-T_{ji\vec{n}}P_{\vec{n}} \right) \Theta_{ij} \nonumber \\
&=&\sum^{J_L-n}\delta \sum_{i=2}^S\sum_{j=2,j\ne i}^S\left(T_{ij\vec{n}+\vec{e}_i-\vec{e}_j}P_{\vec{n}+\vec{e}_i-\vec{e}_j}-T_{ji\vec{n}}P_{\vec{n}} \right) \Theta_{ij} \nonumber \\
&&+\sum^{J_L-n}\delta \sum_{j=2}^S\left(T_{1j\vec{n}+\vec{e}_1-\vec{e}_j}P_{\vec{n}+\vec{e}_1-\vec{e}_j}-T_{j1\vec{n}}P_{\vec{n}} \right) \Theta_{1j} \nonumber \\
&&+\sum^{J_L-n}\delta \sum_{i=2}^S\left(T_{i1\vec{n}+\vec{e}_i-\vec{e}_1}P_{\vec{n}+\vec{e}_i-\vec{e}_1}-T_{1i\vec{n}}P_{\vec{n}} \right) \Theta_{i1}. 
\eeqa{dpndtmarg}
By inspection, the first term is identically zero and the remaining terms generate the right-hand side of Eq.~\ref{margprocessasymm}, namely  
\beqa
\sum^{J_L-n}\delta \sum_{j=2}^S T_{1j\vec{n}+\vec{e}_1-\vec{e}_j}P_{\vec{n}+\vec{e}_1-\vec{e}_j}\Theta_{1j} &=& r_{n+1}\Theta(J_L-(n+1))P_{n+1}, \nonumber \\
\sum^{J_L-n}\delta \sum_{j=2}^S T_{j1\vec{n}}P_{\vec{n}} \Theta_{1j} &=& g_n\Theta(J_L-(n+1))P_n,\nonumber \\
\sum^{J_L-n}\delta \sum_{i=2}^S T_{i1\vec{n}+\vec{e}_i-\vec{e}_1}P_{\vec{n}+\vec{e}_i-\vec{e}_1} \Theta_{i1}&=& g_{n-1}\Theta(n-1)P_{n-1}, \nonumber \\
\sum^{J_L-n}\delta \sum_{i=2}^S T_{1i\vec{n}}P_{\vec{n}} \Theta_{i1}&=&r_n\Theta(n-1)P_n.
\eeqa{equate4and12}
To provide an illustration, let  $J_L=4$, $S=3$, and $n=1$.  Then, 
\beqa
&&\sum^{J_L-n}\delta \sum_{j=2,j\ne i}^ST_{1j\vec{n}+\vec{e}_1-\vec{e}_j}P_{\vec{n}+\vec{e}_1-\vec{e}_j}\Theta_{1j} \nonumber \\
&=&\left(P_{(2,0,2)}(T_{12(2,0,2)}+T_{13(2,0,2)})+P_{(2,1,1)}(T_{12(2,1,1)}+T_{13(2,1,1)})\right.\nonumber \\
&&\left.+P_{(2,2,0)}(T_{12(2,2,0)}+T_{13(2,2,0)})\right)\Theta(J_L-(n+1)) \nonumber \\
&=&r_{n+1}\Theta(J_L-(n+1))(P_{(2,0,2)}+P_{(2,1,1)}+P_{(2,2,0)}) \nonumber \\
&=&r_{n+1}\Theta(J_L-(n+1))P_{n+1},
\eeqa{sumexample}
where we have used the definitions of $\Theta_{ij}$ from Eq.~\ref{Theta}, $r_n$ from Eq.~\ref{dn}, and $P_n$ from Eq.~\ref{pnexampleasymm}.

We construct an approximation to the multivariate sampling distribution of a nearly neutral community, $P^{{\rm nn}*}_{\vec{n}}$, by following the subsample approach of~\citet{Etienne:2005p3829} and~\citet{Etienne:2007p5640} that centers on the identity.
\beq
P^{{\rm nn}*}_{\vec{n}}\,=\,P_n^{{\rm nn}*}\prod_{f=2}^{S-1}P_{n_f|n,n_2,\dots,n_{f-1}}^{{\rm nn}*}.
\eeq{PnvecNNAppB}
Assuming that $P_n$, $P_{n_2|n}$, $\dots$, $P_{n_{f-1}|n,n_2,\dots,n_{f-2}}$ are nonzero and stationary, we argue that conditional marginal dynamics for $P_{n_f|n,n_2,\dots,n_{f-1}}$ are approximated by the master equation
\beqa
\frac{dP_{n_f|n,n_2,\dots,n_{f-1}}}{d\tau}&=&g_{n_f-1|n,n_2,\dots,n_{f-1}}\Theta(n_f-1)P_{n_f-1|n,n_2,\dots,n_{f-1}} \nonumber \\
&&+r_{n_f+1|n,n_2,\dots,n_{f-1}}\Theta(J_L-(n_f+1))P_{n_f+1|n,n_2,\dots,n_{f-1}} \nonumber \\
&&-\left(g_{n_f|n,n_2,\dots,n_{f-1}}\Theta(J_L-(n_f+1)) \right. \nonumber \\
&&\hskip0.26cm \left.+r_{n_f|n,n_2,\dots,n_{f-1}}\Theta(n_f-1)\right)P_{n_f|n,n_2,\dots,n_{f-1}},
\eeqa{margprocesssymm}
where
\beqa
g_{n_f|n,n_2,\dots,n_{f-1}}\hskip-0.3cm&\equiv&\hskip-0.3cm\sum_{i=f+1}^S T_{if(n,n_2,\dots,n_f,\dots,n_S)} \nonumber \\
&=&\hskip-0.3cm\frac{J_L-\tilde{n}_f}{J_L}\left((1-m_o)\frac{n_f}{J_L+n(w-1)-1}+m_o\frac{x_f}{1+x(w-1)}\right)\hskip-0.1cm, \nonumber \\
r_{n_f|n,n_2,\dots,n_{f-1}}\hskip-0.3cm&\equiv&\hskip-0.3cm\sum_{i=f+1}^S T_{fi(n,n_2,\dots,n_f,\dots,n_S)} \nonumber \\
&=&\hskip-0.3cm\frac{n_f}{J_L}\left((1-m_o)\frac{J_L-\tilde{n}_f}{J_L+n(w-1)-1}+m_o\frac{1-\tilde{x}_f}{1+x(w-1)}\right)\hskip-0.1cm, \nonumber \\
\eeqa{bndnf}
and
\beqa
\tilde{n}_f&=&n+\sum_{k=2}^fn_k, \nonumber \\
\tilde{x}_f&=&x+\sum_{k=2}^fx_k. 
\eeqa{defNN}
Eq.~\ref{margprocesssymm} can be derived from the multivariate dynamics of Eq.~\ref{evolve1} under the approximation that stochastic variables $(N(\tau),\allowbreak N_2(\tau),\allowbreak \dots,\allowbreak N_{f-1}(\tau))=(n,\allowbreak n_2,\allowbreak \dots,\allowbreak n_{f-1})$ are fixed in time such that $T_{ij\vec{n}}=0$ for $i,j<f$.  In this scenario, the summations on the right-hand side of Eq.~\ref{evolve1} may begin at $f$
\beq
\frac{dP_{\vec{n}}}{d\tau}\,=\,\sum_{i=f}^S\sum_{j=f,j\ne i}^S\left(T_{ij\vec{n}+\vec{e}_i-\vec{e}_j}P_{\vec{n}+\vec{e}_i-\vec{e}_j}-T_{ji\vec{n}}P_{\vec{n}} \right) \Theta_{ij}.
\eeq{evolve1f}
Given the identity
\beq
P_{\vec{n}}\,=\,P_n P_{n_2|n} \cdots P_{n_{f-1}|n,n_2,\dots,n_{f-2}} P_{n_f,\dots,n_S|n,n_2,\dots,n_{f-1}},
\eeq{pnid}
the stationary factor $P_n P_{n_2|n} \cdots P_{n_{f-1}|n,n_2,\dots,n_{f-2}} $ cancels from both sides of Eq.~\ref{evolve1f} to yield
\beqa
&&\frac{dP_{\vec{n}_f|n,n_2,\dots,n_{f-1}}}{d\tau} \nonumber \\
&=&\sum_{i=f}^S\sum_{j=f,j\ne i}^S\left(T_{ij\vec{n}+\vec{e}_i-\vec{e}_j}P_{\vec{n}_f+\vec{e}_{fi}-\vec{e}_{fj}|n,n_2,\dots,n_{f-1}}-T_{ji\vec{n}}P_{\vec{n}_f|n,n_2,\dots,n_{f-1}} \right) \Theta_{ij}, \nonumber \\
\eeqa{evolve1ff}
where $\vec{n}_f\equiv(n_f,\dots,n_S)$ and $e_{fi}$ is an $(S-f+1)$--dimensional unit vector along the $i$th--direction.  Now let 
\beq
\sum^{J_L-\cdots-n_f}\delta\,\equiv\,\sum_{n_{f+1}=0}^{J_L-n-n_2-\cdots-n_f} \dots \sum_{n_S=0}^{J_L-n-n_2-\cdots-n_f}\delta(J_L-n-n_2-\dots-n_S),
\eeq{defsumdeltasymm}
so that 
\beq
P_{n_f|n,n_2,\dots,n_{f-1}}\,=\,\sum^{J_L-\cdots-n_f}\delta P_{\vec{n}_f|n,n_2,\dots,n_{f-1}}.
\eeq{pnexamplesymm}
Applying Eq.~\ref{defsumdeltasymm} to both sides of Eq.~\ref{evolve1ff}, we obtain
\beqa
&&\frac{dP_{n_f|n,n_2,\dots,n_{f-1}}}{d\tau} \nonumber \\
&=&\sum^{J_L-\cdots-n_f}\delta \sum_{i=f+1}^S\sum_{j=f+1,j\ne i}^S\left(T_{ij\vec{n}+\vec{e}_i-\vec{e}_j}P_{\vec{n}_f+\vec{e}_{fi}-\vec{e}_{fj}|n,n_2,\dots,n_{f-1}}\right. \nonumber \\
&&\hskip8cm\left.-T_{ji\vec{n}}P_{\vec{n}_f|n,n_2,\dots,n_{f-1}} \right) \Theta_{ij} \nonumber \\
&&+\sum^{J_L-\cdots-n_f}\delta \sum_{j=f+1}^S\left(T_{fj\vec{n}+\vec{e}_f-\vec{e}_j}P_{\vec{n}_f+\vec{e}_{ff}-\vec{e}_{fj}|n,n_2,\dots,n_{f-1}}\right. \nonumber \\
&&\hskip8cm\left.-T_{jf\vec{n}}P_{\vec{n}_f|n,n_2,\dots,n_{f-1}} \right) \Theta_{fj} \nonumber \\
&&+\sum^{J_L-\cdots-n_f}\delta \sum_{i=f+1}^S \left(T_{if\vec{n}+\vec{e}_i-\vec{e}_f}P_{\vec{n}_f+\vec{e}_{fi}-\vec{e}_{ff}|n,n_2,\dots,n_{f-1}}\right. \nonumber \\
&&\hskip8cm\left.-T_{fi\vec{n}}P_{\vec{n}_f|n,n_2,\dots,n_{f-1}} \right) \Theta_{if}. \nonumber \\
\eeqa{dpndtcondmarg}
By inspection, the first term is identically zero and the remaining terms generate the right-hand side of Eq.~\ref{margprocesssymm}, namely  
\beqa
&&\sum^{J_L-\cdots-n_f}\delta \sum_{j=f+1}^S T_{fj\vec{n}+\vec{e}_f-\vec{e}_j}P_{\vec{n}_f+\vec{e}_{ff}-\vec{e}_{fj}|n,n_2,\dots,n_{f-1}} \Theta_{fj} \nonumber \\
&&\hskip4cm=r_{n_f+1|n,n_2,\dots,n_{f-1}}\Theta(J_L-(n_f+1))P_{n_f+1|n,n_2,\dots,n_{f-1}}, \nonumber \\
&&\sum^{J_L-\cdots-n_f}\delta \sum_{j=f+1}^S T_{jf\vec{n}}P_{\vec{n}_f|n,n_2,\dots,n_{f-1}} \Theta_{fj} \nonumber \\
&&\hskip4cm= g_{n_f|n,n_2,\dots,n_{f-1}}\Theta(J_L-(n_f+1))P_{n_f|n,n_2,\dots,n_{f-1}},\nonumber \\
&&\sum^{J_L-\cdots-n_f}\delta \sum_{i=f+1}^S T_{if\vec{n}+\vec{e}_i-\vec{e}_f}P_{\vec{n}_f+\vec{e}_{fi}-\vec{e}_{ff}|n,n_2,\dots,n_{f-1}} \Theta_{if} \nonumber \\
&&\hskip4cm= g_{n_f-1|n,n_2,\dots,n_{f-1}}\Theta(n_f-1)P_{n_f-1|n,n_2,\dots,n_{f-1}}, \nonumber \\
&&\sum^{J_L-\cdots-n_f}\delta \sum_{i=f+1}^S T_{fi\vec{n}}P_{\vec{n}_f|n,n_2,\dots,n_{f-1}} \Theta_{if} \nonumber \\
&&\hskip4cm= r_{n_f|n,n_2,\dots,n_{f-1}}\Theta(n_f-1)P_{n_f|n,n_2,\dots,n_{f-1}}.
\eeqa{equate4and23}
The stationary distribution of Eq.~\ref{margprocesssymm} is a P\'olya distribution~\citep{Johnson:1992tUDD}
\beqa
&P_{n_f|n,n_2,\dots,n_{f-1}}^{{\rm nn}*}& \nonumber \\
&=&\hskip-1.2cm N_{Pf} \binom{J_L-\tilde{n}_{f-1}}{n_f} \frac{{\rm B}\left(\phi_{on} x_f+n_f,J_L-\tilde{n}_{f-1}+\phi_{on}\left(1-\tilde{x}_f\right)-n_f\right)}{{\rm B}\left(\phi_{on} x_f,J_L-\tilde{n}_{f-1}+\phi_{on}\left(1-\tilde{x}_f\right)\right)}, \nonumber \\
\eeqa{Pnf}
where
\beq
N_{Pf}^{-1}\,=\, \frac{{\rm B}\left(J_L-\tilde{n}_{f-1}+\phi_{on}\left(1-\tilde{x}_{f-1}\right),\phi_{on}\left(1-\tilde{x}_f\right)\right)}{{\rm B}\left(\phi_{on}\left(1-\tilde{x}_{f-1}\right),J_L-\tilde{n}_{f-1}+\phi_{on}\left(1-\tilde{x}_f\right)\right)},
\eeq{NPf}
and
\beq
\phi_{on}\,=\,I_o\frac{1+n(w-1)/(J_L-1)}{1+x(w-1)}.
\eeq{defphi2}
Plugging Eqs.~\ref{PnNN} and \ref{Pnf} into \ref{PnvecNNAppB}, we obtain the approximate sampling distribution of Eq.~\ref{PnvecNN}.  To validate Eq.~\ref{PnvecNN}, we demonstrate approximate detailed balance, as defined by Eq.~\ref{detailedbalance}, for large--$J_L$ nearly neutral communities where $S,w-1<<\sum_{l=1}^Sw_ln_l-1=J_L-1+n(w-1)$ such that $\epsilon\equiv (w-1)/(J_L-1+n(w-1))$ is a small number.  For $i,j\ge 2$, detailed balance is exact.  But for $i=1$ and $j\ge 2$, we have
\beqa
\frac{P^{{\rm nn}*}_{\vec{n}}}{P^{{\rm nn}*}_{\vec{n}+\vec{e}_1-\vec{e}_j}}&=&\frac{n+1}{n_j}\frac{1}{w}\frac{1-m_o}{1-m}\frac{n_j-1+\phi_{on}x_j}{n+\phi_n x} \nonumber \\
&&\times\frac{((1-x)\phi_{on+1})_{J_L-n-1}}{((1-x)\phi_{on})_{J_L-n-1}}\prod_{k=2}^S\frac{\left(\phi_{on}x_k\right)_{n_k-\delta_{jk}}}{\left(\phi_{on+1}x_k\right)_{n_k-\delta_{jk}}},
\eeqa{gensamptestaagain}
where
\beq
\phi_{n}\,=\,I\frac{1+n(w-1)/(J_L-1)}{1+x(w-1)}.
\eeq{defphi3}
Given $(a(1+\epsilon))_n\sim(a)_n+{\cal O}(\epsilon)$, we find
\beq
\frac{P^{{\rm nn}*}_{\vec{n}}}{P^{{\rm nn}*}_{\vec{n}+\vec{e}_1-\vec{e}_j}}\,=\,\frac{T_{1j\vec{n}+\vec{e}_1-\vec{e}_j}}{T_{j1\vec{n}}}+{\cal O}(S\epsilon).
\eeq{gensamptestbagain}
The case of $i\ge 2$ and $j=1$ is similar.


\makeatletter
\@addtoreset{equation}{subsubsection}
\makeatother
\renewcommand{\theequation}{C.\arabic{subsection}.\arabic{subsubsection}.\arabic{equation}}

\section{Asymptotic Expansions for Hypergeometric Functions}
\label{app}

Calculating Eqs.~\ref{PnNN}, \ref{avgn}, \ref{avgn2}, \ref{PnvecNN}, \ref{PnNNM}, \ref{PnvecNNM}, and \ref{PnhatvecMlimit} for large communities requires computationally intensive evaluations of hypergeometric functions.  To address this problem, one of us (N.M.T.) developed previously unknown asymptotic expansions.  All required expansions are summarized here.  Relevant details can be found in~\cite{Abra65,Luke:1969:SFA,Wong:2001:AAI,Gil:2007:NSF}.

\subsection{Expanding ${}_3F_2(1-J_L,1+\lambda,2;2-\xi,1;\eta)$}
\label{otherform}
Using the reduction formula
\begin{equation}\label{oth1}
{}_3F_2(a,b,2;c,1;z)=\frac{abz}{c}{}_2F_1(a+1,b+1;c+1;z)+{}_2F_1(a,b;c;z),
\end{equation}
this case can be expanded with the methods of \S\ref{secondsec}.

\subsection{Expanding ${}_2F_1(\alpha-J_L,\alpha+\lambda;\alpha+1-\xi;\eta)$}\label{secondsec}
\subsubsection{Notation}\label{not}
We write
\begin{equation}\label{n_1}
a=\alpha-J_L,\quad b=\alpha+\beta+\mu J_L,\quad c=\alpha+\gamma+\rho J_L,
\end{equation}
with $\alpha=0,1,2$ and $J_L$ a positive integer.  In terms of $w$, $m$, $x$, and $m_o$ we have
\begin{equation}\label{n_2}
\beta=-\frac{mx}{1-m+x(w-1)},\quad\mu=-\beta,
\end{equation}
and
\begin{equation}\label{n_3}
\gamma=\frac{1-xwm_o+x(w-1)}{1-wm_o+x(w-1)},\quad \rho =-\gamma.
\end{equation}
The asymptotic behaviour will be considered  of the Gauss hypergeometric function
\begin{equation}\label{n_4}
F={}_2F_1(a,b;c;\eta),
\end{equation}
for large--$J_L$, where
\begin{equation}\label{n_5}
\eta=w\frac{1-m+x(w-1)}{1-wm_o+x(w-1)},
\end{equation}
and
\begin{equation}\label{n_6}
w\in(0,\infty),\quad x,m,m_o\in(0,1).
\end{equation}

\subsubsection{The neutral case: $w=1,$ $m=m_o$}\label{r1}
In this case
\begin{equation}\label{r_}
\eta=1,\quad \mu=\frac{mx}{1-m},\quad \rho=-\frac{1-mx}{1-m}.
\end{equation}
The exact relation
\begin{equation}\label{r1_2}
{}_2F_1(-n,b;c;1)=\frac{(c-b)_n}{(c)_n}=\frac{\Gamma(c)\Gamma(c-b+n)}{\Gamma(c+n)\Gamma(c-b)},\quad n=0,1,2,\ldots,
\end{equation}
can be used, together with the asymptotic estimate of the ratio of gamma functions
\begin{equation}\label{r1_3}
\frac{\Gamma(x+n)}{\Gamma(y+n)}=n^{x-y}\left(1+\bigO(1/n)\right),\quad n\to\infty.
\end{equation}

\subsubsection{Critical values}\label{crit}
Considered as functions of $w$, $\mu$ and $\rho$ become unbounded at $w=w_{c_\mu}$ and 
$w=w_{c_\rho}$, respectively, where
\begin{equation}\label{cv1}
w_{c_\mu}=\frac{m+x-1}{x},\quad w_{c_\rho}=\frac{1-x}{m_o-x}.
\end{equation}

\subsubsection*{The case $w\to w_{c_\mu}$}
\label{rcmu}
In this case $\eta$ becomes small, $b$ becomes unbounded, but the product $b\eta$ remains finite. The $k$th term of the standard power series of $F$ becomes (see also \eqref{r1_3})
\begin{equation}\label{cv2}
\frac{(a)_k(b)_k}{k! (c)_k}\eta^k\sim
\frac{(a)_k}{k! (c_0)_k}z^k,
\end{equation}
with 
\begin{equation}\label{cv3}
z=\lim_{w\to w_{c_\mu}} b\eta=u+vJ_L,\quad c_0=\lim_{w\to w_{c_\mu}}c=\gamma_0+\rho_0J_L,
\end{equation}
where 
\begin{equation}\label{cv4}
u=-\frac{mx(m+x-1)}{mx-m_o(m+x-1)},\quad v=-u,
\end{equation}
and
\begin{equation}\label{cv5}
\gamma_0=\frac{x(m(1-m_o)+m_o(1-x))}{mx-m_o(m+x-1)},\quad
\rho_0=-\gamma_0.
\end{equation}
It follows that $F$ approaches a confluent hypergeometric function:
\begin{equation}\label{cv6}
{}_2F_1(a,b;c;\eta)\to{}_1F_1(a;c_0;z).
\end{equation}
Further action is needed to obtain an asymptotic approximation of the ${}_1F_1$--function.

\subsubsection*{The case $w\to w_{c_\rho}$}
\label{rcnu}
In this case $\eta$ and $c$ become unbounded, but the ratio $\eta/c$ remains finite. The $k$th term of the standard power series of $F$ becomes 
\begin{equation}\label{cv7}
\frac{(a)_k(b)_k}{k! (c)_k}\eta^k\sim
\frac{(a)_k(b_0)_k}{k!\,z^k},
\end{equation}
with 
\begin{equation}\label{cv8}
z=\lim_{w\to w_{c_\rho}} c/\eta=u+vJ_L,\quad b_0=\lim_{w\to w_{c_\rho}}b=\beta_0+\mu_0J_L,
\end{equation}
where 
\begin{equation}\label{cv9}
u=\frac{m_o(m_o-x)(1-x)}{mx-m_o(m+x-1)},\quad v=-u,
\end{equation}
and
\begin{equation}\label{cv10}
\beta_0=-\frac{mx(m_o-x)}{mx-m_o(m+x-1)},\quad
\mu_0=-\beta_0.
\end{equation}
It follows that $F$ approaches a ${}_2F_0$ hypergeometric function
\begin{equation}\label{cv11}
{}_2F_1(a,b;c;\eta)\to{}_2F_0(a,b_0;-;1/z)=\sum_{k=0}^{-a}\frac{(a)_k(b_0)_k}{k! \,z^k},
\end{equation}
because $a$ is  a negative integer. This function can be expressed in terms of the Kummer $U$--function
\begin{equation}\label{cv12}
{}_2F_0(a,b_0;-;1/z)=(-z)^aU(a,1+a-b_0,-z).
\end{equation}
Further action is needed to obtain an asymptotic approximation of the $U$--function.

\subsubsection{Expansion A}\label{simpsad}
An integral representation is
\begin{equation}\label{ss1}
{}_2F_1(a,b;c;\eta)=\frac{\Gamma(c)}{\Gamma(b)\Gamma(c-b)}\int_0^1t^{b-1}(1-t)^{c-b-1}(1-t\eta)^{-a}\,dt,
\end{equation}
valid for $c> b>0, \eta<1$. This integral can be used when
$\rho>\mu>0, \eta<1$.

As an example, consider 
\begin{equation}\label{ss3}
r=3, \quad m=\tfrac12, \quad m_o=\tfrac12, \quad x=\tfrac13.
\end{equation}
This gives
\begin{equation}\label{ss4}
b=\alpha+\tfrac1{11}(J_L-1), \quad c=\alpha+5(J_L-1), \quad \mu=\tfrac1{11},\quad \rho= 5,\quad \eta=-11.
\end{equation}

In this case the integrand becomes small at $t=0$ and $t=1$, and there is a maximum of the integrand at $t=t_1$, with $t_1\in(0,1)$. This point gives the main contribution.

Write  \eqref{ss1} as
\begin{equation}\label{ss5}
{}_2F_1(a,b;c;\eta)=\frac{\Gamma(c)}{\Gamma(b)\Gamma(c-b)}
\int_0^1t^{\alpha+\beta-1}(1-t)^{\gamma-\beta-1}(1-t\eta)^{-\alpha}e^{-J_L\phi(t)}\,dt,
\end{equation}
where
\begin{equation}\label{ss6}
\phi(t)=-\mu\ln(t)-(\rho-\mu)\ln(1-t)-\ln(1-t\eta).
\end{equation}
The saddle points $t_0$ and $t_1$ are the zeros of $\phi^\prime(t)$. For the example \eqref{ss3} this gives
\begin{equation}\label{ss7}
t_0=-0.01169\cdots,\quad t_1 =0.1178\cdots,
\end{equation}
and
\begin{equation}\label{ss8}
\phi(t_1) =-0.02136\cdots, \quad \phi^{\prime\prime}(t_1)=35.83\cdots\,.
\end{equation}

An asymptotic approximation follows from the substitution
\begin{equation}\label{ss9}
\phi(t)-\phi(t_1) =\tfrac12\phi^{\prime\prime}(t_1)s^2,\quad \sign(t-t_1)=\sign(s),
\end{equation}
which gives
\begin{equation}\label{ss10}
{}_2F_1(a,b;c;\eta)=\frac{\Gamma(c)}{\Gamma(b)\Gamma(c-b)}
e^{-J_L\phi(t_1)}\int_{-\infty}^\infty f(s) e^{-\tfrac12J_L\phi^{\prime\prime}(t_1)s^2}\,ds,
\end{equation}
where
\begin{equation}\label{ss11}
f(s)=t^{\alpha+\beta-1}(1-t)^{\gamma-\beta-1}(1-t\eta)^{-\alpha}\frac{dt}{ds}.
\end{equation}
Because locally at $t=t_1$ (or $s=0$), $t=t_1+s+\bigO(s^2)$, we have $dt/ds=1$ at $s=0$, and 
\begin{equation}\label{ss12}
f(0)=t_1^{\alpha+\beta-1}(1-t_1)^{\gamma-\beta-1}(1-t_1\eta)^{-\alpha}.
\end{equation}
This gives the first order approximation
\begin{equation}\label{ss13}
{}_2F_1(a,b;c;\eta)\sim\frac{\Gamma(c)}{\Gamma(b)\Gamma(c-b)}
e^{-J_L\phi(t_1)}f(0)\int_{-\infty}^\infty e^{-\tfrac12J_L\phi^{\prime\prime}(t_1)s^2}\,ds,
\end{equation}
that is
\begin{equation}\label{ss14}
{}_2F_1(a,b;c;\eta)\sim\frac{\Gamma(c)}{\Gamma(b)\Gamma(c-b)}
e^{-J_L\phi(t_1)}f(0)\sqrt{\frac{2\pi}{J_L\phi^{\prime\prime}(t_1)}},\quad J_L\to\infty.
\end{equation}

\subsubsection{Expansion B}\label{othsad}
Another integral representation is
\begin{equation}\label{os1}
{}_2F_1(a,b;c;\eta)=\frac{\Gamma(1+b-c)}{\Gamma(b)\Gamma(1-c)}\int_0^{\infty}t^{b-1}(t+1)^{c-b-1}(1+t\eta)^{-a}\,dt,
\end{equation}
which is only valid for $a=0,-1,-2,\ldots$ and $c<a+1$. It can be verified by expanding $(1+t\eta)^{-a}$ in powers of $\eta$.

We have $\mu>0$ and $\rho<-1$, and because (see \eqref{n_2}, \eqref{n_3} and \eqref{n_5})
\begin{equation}\label{os2}
\eta=-\frac{mx}{(1-m_ox)}\,\frac{\rho}{\mu},
\end{equation}
we see that $\eta\ge0$.

As an example, consider
\begin{equation}\label{os3}
r=\tfrac13, \quad m=\tfrac12, \quad m_o=\tfrac12, \quad x=\tfrac13.
\end{equation}
This gives
\begin{equation}\label{os4}
b=\alpha+3(J_L-1), \quad c=\alpha+\tfrac{15}{13}(1-J_L), \quad \mu=3,\quad \rho= -\tfrac{15}{13},\quad \eta=\tfrac{1}{13}.
\end{equation}

Write  \eqref{os1} as
\begin{equation}\label{os5}
{}_2F_1(a,b;c;\eta)=\frac{\Gamma(1+b-c)}{\Gamma(b)\Gamma(1-c)}\int_0^{\infty}
t^{\alpha+\beta-1}(t+1)^{\gamma-\beta-1}(1+t\eta)^{-\alpha}e^{-J_L\psi(t)}\,dt,
\end{equation}
where
\begin{equation}\label{os6}
\psi(t)=-\mu\ln(t)-(\rho-\mu)\ln(t+1)-\ln(1+t\eta).
\end{equation}
The saddle points $t_0$ and $t_1$ are  for the example \eqref{os3}
\begin{equation}\label{os7}
t_0=-74.89\cdots,\quad t_1 =3.385\cdots,
\end{equation}
and
\begin{equation}\label{os8}
\psi(t_1) =2.251\cdots, \quad \psi^{\prime\prime}(t_1)=0.04951\cdots\,.
\end{equation}

An asymptotic approximation follows from the substitution
\begin{equation}\label{os9}
\psi(t)-\psi(t_1) =\tfrac12\psi^{\prime\prime}(t_1)s^2,\quad \sign(t-t_1)=\sign(s),
\end{equation}
which gives
\begin{equation}\label{os10}
{}_2F_1(a,b;c;\eta)=\frac{\Gamma(1+b-c)}{\Gamma(b)\Gamma(1-c)}
e^{-J_L\psi(t_1)}\int_{-\infty}^\infty g(s) e^{-\tfrac12J_L\psi^{\prime\prime}(t_1)s^2}\,ds,
\end{equation}
where
\begin{equation}\label{os11}
g(s)=t^{\alpha+\beta-1}(1+t)^{\gamma-\beta-1}(1+t\eta)^{-\alpha}\frac{dt}{ds}.
\end{equation}
Because locally at $t=t_1$ (or $s=0$), $t=t_1+s+\bigO(s^2)$, we have $dt/ds=1$ at $s=0$, and 
\begin{equation}\label{os12}
g(0)=t_1^{\alpha+\beta-1}(1+t_1)^{\gamma-\beta-1}(1+t_1\eta)^{-\alpha}.
\end{equation}
This gives the first order approximation
\begin{equation}\label{os13}
{}_2F_1(a,b;c;\eta)\sim\frac{\Gamma(1+b-c)}{\Gamma(b)\Gamma(1-c)}
e^{-J_L\psi(t_1)}g(0)\int_{-\infty}^\infty e^{-\tfrac12J_L\psi^{\prime\prime}(t_1)s^2}\,ds,
\end{equation}
that is
\begin{equation}\label{os14}
{}_2F_1(a,b;c;\eta)\sim\frac{\Gamma(1+b-c)}{\Gamma(b)\Gamma(1-c)}
e^{-J_L\psi(t_1)}g(0)\sqrt{\frac{2\pi}{J_L\psi^{\prime\prime}(t_1)}},\quad J_L\to\infty.
\end{equation}

\subsubsection{Expansion C}\label{finalsad}

If $\mu<\rho<-1$ and $\eta<0$, apply the transformation 
\begin{equation}\label{rem2}
{}_2F_1(a,b;c;\eta)=(1-\eta^\prime)^a{}_2F_1(a,b^\prime;c;\eta^\prime),
\end{equation}
where 
\begin{equation}\label{rem3}
b^{\prime}=c-b=\beta^\prime+\mu^\prime J_L,\quad \beta^\prime=\gamma-\beta,\quad \mu^\prime=\rho-\mu, \quad \eta^\prime=\frac{\eta}{\eta-1}.
\end{equation}
Now,
\begin{equation}\label{rem4}
\mu^\prime>0,\quad \rho<-1,\quad \eta^\prime>0,
\end{equation}
and it follows that Expansion B, \S \ref{othsad}, applies to the Gauss function on the right-hand side of \eqref{rem2}.

\subsubsection{General cases for all non-critical values}\label{remaining}

\begin{enumerate}

\item
$w_{c_\mu},w_{c_\rho}<0$
\subitem
For all $w>0$, we have $\mu>0$, $\rho<-1$, and $\eta>0$, so use Expansion B, \S\ref{othsad}.

\item
$w_{c_\mu}>0$, $w_{c_\rho}<0$
\subitem
For all $w_{c_\mu}>w>0$, we have $\mu<-1$, $\rho<-1$, and $\eta<0$, so use Expansion C, \S\ref{finalsad}.
\subitem
For all $w>w_{c_\mu}$, we have $\mu>0$, $\rho<-1$, and $\eta>0$, so use Expansion B, \S\ref{othsad}.

\item
$w_{c_\mu}<0$, $w_{c_\rho}>0$
\subitem
For all $w_{c_\rho}>w>0$, we have $\mu>0$, $\rho<-1$, and $\eta>0$, so use Expansion B, \S\ref{othsad}.
\subitem
For all $w>w_{c_\rho}$, we have $\rho>\mu>0$ and $\eta<0$, so use Expansion A, \S\ref{simpsad}.

\item
$w_{c_\rho}>w_{c_\mu}>0$
\subitem
For all $w_{c_\mu}>w>0$, we have $\mu<-1$, $\rho<-1$, and $\eta<0$, so use Expansion C, \S\ref{finalsad}.
\subitem
For all $w_{c_\rho}>w>w_{c_\mu}$, we have $\mu>0$, $\rho<-1$, and $\eta>0$, so use Expansion B, \S\ref{simpsad}.
\subitem
For all $w>w_{c_\rho}$, we have $\rho>\mu>0$ and $\eta<0$, so use Expansion A, \S\ref{simpsad}.

\end{enumerate}

\subsection{Expanding ${}_2F_1(1-J_M,1;2-\xi_M;\eta_M)$}\label{F21forF32}
\subsubsection{Notation}
We write
\begin{equation}\label{othf1}
a=1-J_M,\quad b=1, \quad c=\sigma+\tau J_M,
\end{equation}
with
\begin{equation}\label{othf2}
\sigma=1+\frac{1}{1-w\nu_o},\quad \tau=-\frac{1}{1-w\nu_o}.
\end{equation}
The asymptotic behaviour will be considered  of the Gauss hypergeometric function
\begin{equation}\label{othf3M}
F={}_2F_1(a,b;c;\eta_M)
\end{equation}
for large--$J_M$, where
\begin{equation}\label{othf4M}
\eta_M=\frac{ w(1-\nu)}{1 -w\nu_o},
\end{equation}
and
\begin{equation}\label{othf5}
w\in(0,\infty),\quad \nu,\nu_o\in(0,1).
\end{equation}

\subsubsection{The neutral case: $w=1$, $\nu=\nu_o$}\label{othr1}
In this case $\eta_M=1$ and \eqref{r1_2} can be used to get an exact result in terms of gamma functions.

\subsubsection{The critical case $w_{c_{\nu_o}}=1/\nu_o$}\label{othcrit}
In this case we have (see also \S\ref{rcnu})
\begin{equation}\label{othf6}
{}_2F_1(a,b;c;\eta_M)\to{}_2F_0(a,b;-;1/z)=(-z)^aU(a,1+a-b,-z),
\end{equation}
where
\begin{equation}\label{othf7}
z=-\frac{(J_M-1)\nu_o}{1-\nu}.
\end{equation}

\subsubsection{The case $0<w<w_{c_{\nu_o}}$}\label{othgenrsmall}
Use the integral representation
\begin{equation}\label{othf11}
{}_2F_1(a,b;c;\eta_M)=\frac{\Gamma(1+b-c)}{\Gamma(b)\Gamma(1-c)}\int_0^{\infty}
t^{b-1}f(t)e^{-J_M\phi(t)}\,dt,
\end{equation}
where
\begin{equation}\label{othf12}
f(t)=(1+t)^{\sigma-b-1}(1+\eta_M t)^{-1},\quad \phi(t)=-\tau\ln(1+t)-\ln(1+\eta_M t).
\end{equation}
The saddle point $t_0$ follows from solving $\phi^\prime(t)=0$. This gives
\begin{equation}\label{othf13}
\phi^\prime(t)=-\frac{\tau}{1+t}-\frac{\eta_M}{1+\eta_M t}, \quad t_0=-\frac{\tau+\eta_M}{\eta_M(\tau+1)}.
\end{equation}
In terms of $w$ and $\nu$
\begin{equation}\label{othf14}
t_0=-\frac{(1-w\nu_o)(1-w+w\nu)}{\nu_o(1-\nu)w^2}.
\end{equation}

\begin{enumerate}
\item
If $0<w<1/(1-\nu)\equiv w_{c_\nu}$, then the saddle point is negative, and we can substitute $s=\phi(t)$, giving
\begin{equation}\label{othf15}
{}_2F_1(a,b;c;\eta_M)=\frac{\Gamma(1+b-c)}{\Gamma(b)\Gamma(1-c)}\int_0^{\infty}
s^{b-1}g(s) e^{-J_Ms}\,ds,\end{equation}
where
\begin{equation}\label{othf16}
g(s)=f(t)\left(\frac{t}{s}\right)^{b-1}\,\frac{dt}{ds}=\left(\frac{t}{s}\right)^{b-1}\,\frac{f(t)}{\phi^\prime(t)}.
\end{equation}
Apply Watson's lemma by expanding $g(s)=\sum_{k=0}^\infty g_ks^k$ to obtain
\begin{equation}\label{othf17}
{}_2F_1(a,b;c;\eta_M)\sim\frac{\Gamma(1+b-c)}{\Gamma(b)\Gamma(1-c)}\sum_{k=0}^\infty \frac{\Gamma(b+k)\,g_k}{J_M^{b+k}}.
\end{equation}
To compute the coefficients $g_k$ we first expand $t=\sum_{k=1}^\infty t_ks^k$. The coefficients $t_k$ follow from inverting the expansion
\begin{equation}\label{othf18}
s=-\tau\ln(1+t)-\ln(1+\eta_M t)=\sum_{k=1}^\infty s_kt^k,\quad s_1=-\tau-\eta_M.
\end{equation}
This gives 
\begin{equation}\label{othf19}
t_1=-\frac{1}{\tau+\eta_M}=\frac{1-w\nu_o}{1-w+w\nu},
\end{equation}
and for the first  coefficient in the expansion \eqref{othf17} $g_0=g(0)=t_1^b$.
This gives
\begin{equation}\label{othf20}
{}_2F_1(a,b;c;\eta_M)\sim\frac{\Gamma(1+b-c)}{\Gamma(1-c)}\left(\frac{t_1}{J_M}\right)^{b}.
\end{equation}
\item
If $w>w_{c_\nu}$, then $t_0$ is positive, and Laplace's method can be used, as in \S\S\ref{simpsad}, \ref{othsad}. We substitute
\begin{equation}\label{othf21}
\tfrac12\phi^{\prime\prime}(t_0)s^2=\phi(t)-\phi(t_0),\quad \phi^{\prime\prime}(t_0)=\frac{\eta_M^2(\tau+1)^3}{\tau(\eta_M-1)^2},
\end{equation}
and  obtain
\begin{equation}\label{othf22}
{}_2F_1(a,b;c;\eta_M)\sim\frac{\Gamma(1+b-c)}{\Gamma(b)\Gamma(1-c)}\sqrt{\frac{2\pi}{J_M\phi^{\prime\prime}(t_0)}}e^{-J_M\phi(t_0)}t_0^{b-1}f(t_0),
\end{equation}
where $t_0$ is given in \eqref{othf13}.
\item
If $w=w_{c_\nu}$, then $t_0=0$ and Laplace's method on a half-infinite interval can be used.
\end{enumerate}


\subsubsection{The case $w>w_{c_{\nu_o}}$}\label{othgenrlarge}
Use the integral representation
\begin{equation}\label{othf23}
{}_2F_1(a,b;c;\eta_M)=\frac{\Gamma(c)}{\Gamma(b)\Gamma(c-b)}\int_0^1f(t)e^{-J_M\phi(t)}\,dt,
\end{equation}
where
\begin{equation}\label{othf24}
f(t)=t^{b-1}(1-t)^{\sigma-b-1}(1-\eta_M t)^{-1},\quad \phi(t)=-\tau\ln(1-t)-\ln(1-\eta_M t).
\end{equation}
The saddle point $t_0$ follows from solving $\phi^\prime(t)=0$. This gives
\begin{equation}\label{othf25}
\phi^\prime(t)=\frac{\tau}{1-t}+\frac{\eta_M}{1-\eta_M t}, \quad t_0=\frac{\tau+\eta_M}{\eta_M(\tau+1)}.
\end{equation}
In terms of $w$ and $\nu$
\begin{equation}\label{othf26}
t_0=\frac{(1-w\nu_o)(1-w+w\nu)}{\nu_o(1-\nu)w^2}.
\end{equation}

\begin{enumerate}
\item
If $w<w_{c_\nu}$, then $t_0<0$ and Watson's lemma should be used. 
The result is
\begin{equation}\label{othf27}
{}_2F_1(a,b;c;\eta_M)\sim\frac{\Gamma(c)}{\Gamma(c-b)}\left(\frac{t_1}{J_M}\right)^{b},\quad t_1=\frac{1}{\eta_M+\tau}.\end{equation}
\item
If $w>w_{c_\nu}$, then the saddle point $t_0$ is always inside the interval $(0,1)$, with $t_0\to1$ if $w\to\infty$. Laplace's method should be used. This gives
\begin{equation}\label{othf28}
{}_2F_1(a,b;c;\eta_M)\sim\frac{\Gamma(c)}{\Gamma(b)\Gamma(c-b)}\sqrt{\frac{2\pi}{J_M\phi^{\prime\prime}(t_0)}}e^{-J_M\phi(t_0)}t_0^{b-1}f(t_0),
\end{equation}
where $f, \phi$ and  $t_0$ are given in \eqref{othf24}--\eqref{othf25} and
\begin{equation}\label{othf29}
\phi^{\prime\prime}(t_0)=\frac{\eta_M^2(\tau+1)^3}{\tau(\eta_M-1)^2}.
\end{equation}
\item
If $w=w_{c_\nu}$ then $t_0=0$ and Laplace's method on a half-infinite interval can be used.

\end{enumerate}


\subsection{Expanding ${}_3F_2(1-J_M,1,1;2,2-\xi_M;\eta_M)$}\label{F32case}
\subsubsection{Notation}
We write
\begin{equation}\label{F1}
a=1-J_M, \quad c=\sigma+\tau J_M,
\end{equation}
with 
\begin{equation}\label{F3}
\sigma=1+\frac{1}{1-w\nu_o},\quad \tau=-\frac{1}{1-w\nu_o},
\end{equation}
The asymptotic behaviour will be considered of the hypergeometric function
\begin{equation}\label{othf3MM}
F={}_3F_2(a,1,1;c,2;\eta_M),
\end{equation}
for large--$J_M$, where
\begin{equation}\label{othf4MM}
\eta_M=\frac{ w(1-\nu)}{1 -w\nu_o},
\end{equation}
and
\begin{equation}\label{F5}
r\in(0,\infty),\quad \nu,\nu_o\in(0,1).
\end{equation}
Note that
\begin{equation}\label{F8}
{}_3F_2(a,1,1;c,2;\eta_M)=\sum_{k=0}^\infty\frac{(a)_k(1)_k(1)_k}{k!\,(c)_k(2)_k}\eta_M^k=
\sum_{k=0}^\infty\frac{(a)_k}{(c)_k}\frac{\eta_M^k}{k+1},
\end{equation}
and because $a$ is a negative integer this series terminates at $k=-a$. 

\subsubsection{Representation in terms of a Laplace integral}\label{Lap}
We substitute
\begin{equation}\label{F9}
\frac{1}{k+1}=\int_0^\infty e^{-(k+1)w}\,dw,
\end{equation}
and obtain
\begin{equation}\label{F10}
{}_3F_2(a,1,1;c,2;\eta_M)=\int_0^{\infty}e^{-w}{}_2F_1(a,1;c,z)\,dw,\quad z=\eta_M e^{-w}.
\end{equation}
For this representation we use the results of \S\S\ref{othgenrsmall}, \ref{othgenrlarge} when, with $\eta_M$ replaced by $\eta_M e^{-w}$, the saddle points $t_0$ of \eqref{othf13} and \eqref{othf25} are negative. This gives two cases.

\begin{enumerate}
\item \emph{The case $w<w_{c_{\nu_o}}$ and $w<w_{c_\nu}$.}
For this case we use the results in \eqref{othf15}-\eqref{othf20} with $b=1$. We have
\begin{equation}\label{F11}
{}_2F_1(a,1;c;z)=\frac{1-c}{J_M}g_0+\bigO(1/J_M),
\end{equation}
where $g_0=t_1$ and $t_1=-1/(\tau+\eta_M e^{-w})$. This gives
\begin{equation}\label{F12}
{}_3F_2(a,1,1;c,2;\eta_M)=\frac{c-1}{J_M}\int_0^{\infty}e^{-w}\,\frac{dw}{\tau+\eta_M e^{-w}}+\bigO(1/J_M).
\end{equation}
Evaluating the integral we obtain
\begin{equation}\label{F13}
{}_3F_2(a,1,1;c,2;\eta_M)=\frac{c-1}{\tau J_M}\frac{\ln(1+\eta_M/\tau)}{\eta_M/\tau}+\bigO(1/J_M).
\end{equation}

\item \emph{The case $w>w_{c_{\nu_o}}$ and $w<w_{c_\nu}$.}
In this case we use the results in \eqref{othf23}-\eqref{othf27}, again, with $b=1$ and $\eta_M$ replaced with $\eta_M e^{-w}$. We use Watson's lemma for \eqref{othf21} by substituting $s=\phi(t)$ and obtain
\begin{equation}\label{F14}
{}_2F_1(a,1;c;z)=(c-1)\int_0^\infty e^{-J_Ms}g(s)\,ds,\quad g(s)=\frac{ (1-t)^{\sigma-2}}{1-te^{-w}}\,\frac{dt}{ds}.
\end{equation}
Expanding $g$ at $s=0$ we have $g(s)=g_0+\bigO(s)$, with $g_0=1/(\tau+\eta_M e^{-w})$ and, as in the above case, we derive
\begin{equation}\label{F15}
{}_3F_2(a,1,1;c,2;\eta_M)=\frac{c-1}{\tau J_M}\frac{\ln(1+\eta_M/\tau)}{\eta_M/\tau}+\bigO(1/J_M).
\end{equation}

\end{enumerate}


\subsubsection{Summing the series by integration}\label{SSI}
We replace the Pochhammer symbols in \eqref{F8} by representations in terms of the gamma functions
\begin{equation}\label{F16}
(\alpha)_k=\frac{\Gamma(\alpha+k)}{\Gamma(\alpha)}=(-1)^k\frac{\Gamma(1-\alpha)}{\Gamma(1-\alpha-k)}.\end{equation}
and replace the gamma functions with large positive argument by their asymptotic forms that follow from 
\begin{equation}\label{F17}
\Gamma(az+b)\sim \sqrt{2\pi}\,e^{-az}(az)^{az+b-\frac12},\quad z\to\infty,\quad a>0.
\end{equation}
This gives the remaining two cases.

\begin{enumerate}
\item \emph{The case $w<w_{c_{\nu_o}}$ and $w>w_{c_\nu}$.}
In this case $\tau<-1$ and $ \eta_M>0$. We replace the Pochhammer symbols in \eqref{F8} with the second form in \eqref{F16}. This gives
\begin{equation}\label{F18}
{}_3F_2(a,1,1;c,2;\eta_M)=\sum_{k=0}^\infty F(k),
\end{equation}
where
\begin{equation}\label{F19}
F(k)=\frac{\Gamma(1-a)}{\Gamma(1-c)} \frac{\eta_M^k}{k+1}\frac{\Gamma(1-c-k)}{\Gamma(1-a-k)},
\end{equation}
and we replace the summation in \eqref{F18} by integration, invoking  Euler's summation formula, or the compound trapezoidal rule 
\begin{equation}\label{F20}
{}_3F_2(a,1,1;c,2;\eta_M)\sim\frac{\Gamma(1-a)}{\Gamma(1-c)}\int_0^{J_M} \frac{\eta_M^k}{k+1}\frac{\Gamma(1-c-k)}{\Gamma(1-a-k)}\,dk.
\end{equation}
In Euler's summation formula additional terms occur but in the present case they can be neglected.

We replace the gamma functions by their asymptotic estimates following from \eqref{F17}. This gives
\begin{equation}\label{F21}
{}_3F_2(a,1,1;c,2;\eta_M)\sim \frac{\Gamma(1-a)}{\Gamma(1-c)}\int_0^{J_M} \frac{e^{-\phi(k)}}{k+1}\frac{\sqrt{1-a-k}}{\sqrt{1-c-k}}\,dk,
\end{equation}
where
\begin{equation}\label{F22}
\phi(k)=-k\ln\eta_M-(1-c-k)\ln(1-c-k)+(1-a-k)\ln(1-a-k).
\end{equation}
Then,
\begin{equation}\label{F23}
\phi^\prime(k)=-\ln \eta_M+\ln(1-c-k)-\ln(1-a-k)),
\end{equation}
and the derivative vanishes for $k=k_s$, where
\begin{equation}\label{F24}
k_s=\frac{c-1+\eta_M(1-a)}{1-\eta_M}=k_0+k_1J_M, \quad 
k_0=\frac{\sigma-1}{\eta_M-1},\quad k_1=\frac{\tau+\eta_M}{\eta_M-1}.
\end{equation}
and $0<k_1<1$. The dominant point of the integral in \eqref{F21} is $k=k_s$, and we apply Laplace's method. We substitute
\begin{equation}\label{F25}
\tfrac12\phi^{\prime\prime}(k_s)s^2=\phi(k)-\phi(k_s),\quad \phi^{\prime\prime}(k_s)=-\frac{(\eta_M-1)^2}{J_M\eta_M(\tau+1)}+\bigO(1/J_M^2).
\end{equation}
This gives
\begin{equation}\label{F26}
{}_3F_2(1-J_M,1,1;\sigma+\tau J_M,2;\eta_M)\, \sim \, 
\sqrt{\frac{2\pi}{\phi^{\prime\prime}(k_s)}}\,F(k_s),
\end{equation}
where $F(k)$ is given in \eqref{F19}.  After using \eqref{F17} we obtain
 \begin{equation}\label{F27}
\hskip-2cm {}_3F_2(1-J_M,1,1;\sigma+\tau J_M,2;\eta_M) \nonumber
\end{equation}
\begin{equation}
\sim\sqrt{\frac{2\pi}{\phi^{\prime\prime}(k_s)}}\,\frac{\eta_M-1}{(\eta_M+\tau)J_M}\left(-\frac{\eta_M}{\tau}\right)^{\frac12-c}
\left(\frac{1+\tau}{1-\eta_M}\right)^{a-c}.
\end{equation}

\item \emph{The case $w>w_{c_{\nu_o}}$ and $w>w_{c_\nu}$.}
In this case $\tau>0$ and $ \eta_M<0$.  We use \eqref{F8}, replacing the Pochhammer symbol $(a)_k$  by the second form of \eqref{F16} and $(c)_k$ by the first. This gives
\begin{equation}\label{F28}
{}_3F_2(a,1,1;c,2;\eta_M)=\sum_{k=0}^\infty F(k),
\end{equation}
where
\begin{equation}\label{F29}
F(k)=\frac{\Gamma(J_M)\Gamma(c)(-\eta_M)^k}{(k+1)\Gamma(J_M-k)\Gamma(c+k)}.
\end{equation}
and we replace the summation in \eqref{F28} by integration,\begin{equation}\label{F30}
{}_3F_2(a,1,1;c,2;\eta_M)\sim \int_0^{J_M} F(k)\,dk.
\end{equation}

Applying the asymptotic estimates of gamma functions in \eqref{F17}, we obtain
\begin{equation}\label{F31}
F(k)=\frac{\Gamma(J_M)\Gamma(c)e^{c+J_M}}{2\pi(k+1)}\sqrt{(J_M-k)(c+k)} e^{-\phi(k)},
\end{equation}
where
\begin{equation}\label{F32}
\phi(k)=(c+k)\ln(c+k)+(J_M-k)\ln(J_M-k)-k\ln(-\eta_M).
\end{equation}
We have
\begin{equation}\label{F33}
\phi^\prime(k)=\ln(c+k)-\ln(J_M-k)-\ln(-\eta_M),
\end{equation}
and the saddle point given by 
\begin{equation}\label{F34}
k_s=\frac{c+\eta_M J_M}{\eta_M-1}=k_0+k_1J_M, \quad 
k_0=\frac{\sigma}{\eta_M-1},\quad k_1=\frac{\tau+\eta_M}{\eta_M-1},
\end{equation}
 where, again, $0<k_1<1$. The dominant point of the integral in \eqref{F30} is $k=k_s$, and we apply Laplace's method to this integral.
We substitute
\begin{equation}\label{F35}
\tfrac12\phi^{\prime\prime}(k_s)s^2=\phi(k)-\phi(k_s),\quad \phi^{\prime\prime}(k_s)=-\frac{(\eta_M-1)^2}{J_M\eta_M(\tau+1)}+\bigO(1/J_M^2),
\end{equation}
where $\phi(k)$ is given in \eqref{F32}. This gives
\begin{equation}\label{F36}
{}_3F_2(1-J_M,1,1;\sigma+\tau J_M,2;\eta_M)\, \sim \, 
\sqrt{\frac{2\pi}{\phi^{\prime\prime}(k_s)}}\,F(k_s),
\end{equation}
where $F(k)$ is given in \eqref{F31}.  After using \eqref{F17} we obtain, as in the case above, 
\begin{equation}\label{F37}
\hskip-2cm {}_3F_2(1-J_M,1,1;\sigma+\tau J_M,2;\eta_M) \nonumber
\end{equation}
\begin{equation}
\sim\sqrt{\frac{2\pi}{\phi^{\prime\prime}(k_s)}}\,\frac{\eta_M-1}{(\eta_M+\tau)J_M}\left(-\frac{\eta_M}{\tau}\right)^{\frac12-c}
\left(\frac{1+\tau}{1-\eta_M}\right)^{a-c}.
\end{equation}

\end{enumerate}

\bibliographystyle{model2-names}
\bibliography{Noble_MainText.bib}







\newpage 

\renewcommand{\thefigure}{\arabic{figure}}


\end{document}